\documentclass[twocolumn]{aastex701}

\usepackage{lineno}
\usepackage{hyperref}
\usepackage{amsmath}
\usepackage{amssymb}
\usepackage{graphicx}
\usepackage[utf8]{inputenc}
\usepackage{float}

\DeclareRobustCommand{\ion}[2]{\textup{#1\,\textsc{\lowercase{#2}}}}

\newcommand\fei{\ion{Fe}{i}}
\newcommand\feii{\ion{Fe}{ii}}
\newcommand\srii{\ion{Sr}{ii}}

\newcommand\baii{\ion{Ba}{ii}}
\newcommand\tii{\ion{Ti}{i}}
\newcommand\tiii{\ion{Ti}{ii}}

\newcommand\nai{\ion{Na}{i}}
\newcommand\mgi{\ion{Mg}{i}}
\newcommand\ali{\ion{Al}{i}}
\newcommand\cai{\ion{Ca}{i}}
\newcommand\scii{\ion{Sc}{ii}}
\newcommand\vi{\ion{V}{i}}
\newcommand\vii{\ion{V}{ii}}
\newcommand\cri{\ion{Cr}{i}}
\newcommand\crii{\ion{Cr}{ii}}
\newcommand\mni{\ion{Mn}{i}}
\newcommand\mnii{\ion{Mn}{ii}}
\newcommand\coi{\ion{Co}{i}}
\newcommand\nii{\ion{Ni}{i}}
\newcommand\zni{\ion{Zn}{i}}

\newcommand\sii{\ion{Si}{i}}

\newcommand{\kms}{km\,s$^{-1}$}
\newcommand{\teff}{$T_{\rm eff}$}
\newcommand{\logg}{$\log g$}
\newcommand{\vt}{$\xi_{t}$}

\newcommand{\rproc}{$r$-process}
\newcommand{\sproc}{$s$-process}

\newcommand{\AB}[2]{$\mbox{[#1/#2]}$}
\newcommand{\feh}{\AB{Fe}{H}}

\newcommand{\thestar}{J0010$-$1735}

\newcommand{\rII}{$r$-II}

\newcommand{\vmic}{$v_\textrm{mic}$}
\newcommand{\chandmass}{M$_{\text{CH}}$}

\usepackage[utf8]{inputenc}

\shorttitle{J0010$-$1735: An Accreted MW Halo star with an Unusual Abundance Pattern}
\shortauthors{Hackshaw et al.}
\begin{document}

\title{The $R$-Process Alliance: On J0010$-$1735, an Accreted, Chemically Unusual, Metal-poor Star with $R$-process 
Enhancement \footnote{This paper includes data gathered with the 6.5 meter Magellan Telescopes located at Las Campanas Observatory, Chile.}}                         

\newcommand{\alex}[1]{\textcolor{orange}{(APJ: #1)}}
\newcommand{\alexs}[2]{\textcolor{orange}{(APJ: \sout{#1} #2)}}

\correspondingauthor{Zoe Hackshaw}

\author[0000-0002-3855-3060]{Zoe Hackshaw}
\affiliation{Department of Astronomy, The University of Texas at Austin,
2515 Speedway Boulevard Austin, TX 78712, USA}
\email{zoehackshaw@utexas.edu}

\author[0000-0002-8504-8470]{Rana Ezzeddine}
\affiliation{Department of Astronomy, University of Florida, Gainesville, FL 32611, USA}
\affiliation{Joint Institute for Nuclear Astrophysics - Center for Evolution of the Elements, USA}
\email{rezzeddine@ufl.edu}

\author[0000-0002-3367-2394]{Shivani P. Shah}
\affiliation{North Carolina State University, Department of Physics and Astronomy, Raleigh, NC 27695, USA}
\email{spshah7@ncsu.edu}

\author[0000-0001-6154-8983]{Terese T. Hansen}
\affiliation{Department of Astronomy, Stockholm University, AlbaNova University Center, SE-106 91 Stockholm, Sweden}
\email{thidemannhansen@gmail.com}

\author[0009-0009-9075-3180]{Ruijie Shi}
\affiliation{Department of Astronomy, University of Florida, Gainesville, FL 32611, USA}
\email{ruijie.shi@ufl.edu}

\author{Yangyang Li}
\affiliation{Department of Astronomy, University of Florida, Gainesville, FL 32611, USA}
\email{yangyangli@ufl.edu}
\author{Timothy C. Beers}
\affiliation{Department of Physics, University of Notre Dame, Notre Dame, IN 46556, USA}
\affiliation{Joint Institute for Nuclear Astrophysics - Center for Evolution of the Elements, USA}
\email{tbeers@nd.edu}

\author{Anna Frebel}
\affiliation{Department of Physics \& Kavli Institute for Astrophysics and Space Research, \\ Massachusetts Institute of Technology, Cambridge, MA 02139, USA}
\affiliation{Joint Institute for Nuclear Astrophysics - Center for Evolution of the Elements, USA}
\email{afrebel@mit.edu}

\author{Erika M. Holmbeck}
\affiliation{Lawrence Livermore National Laboratory, 7000 East Avenue, Livermore, CA 94550, USA}
\email{erikaholmbeck@gmail.com}
\author{Alexander P. Ji}
\affiliation{Department of Astronomy \& Astrophysics, University of Chicago, 5640 S. Ellis Avenue, Chicago, IL 60637, USA}
\affiliation{Kavli Institute for Cosmological Physics, University of Chicago, 5640 S Ellis Avenue, Chicago, IL 60637, USA}
\email{alexji@uchicago.edu}
\author{Jeremy Kowkabany}
\affiliation{Department of Astronomy, University of Florida, Gainesville, FL 32611, USA}
\email{jk21s@my.fsu.edu}
\author[0000-0003-4479-1265]{Vinicius M. Placco}
\affiliation{NSF NOIRLab, Tucson, AZ 85719, USA}
\email{vinicius.placco@noirlab.edu}
\author[0000-0001-5107-8930]{Ian U. Roederer}
\affiliation{North Carolina State University, Department of Physics and Astronomy, Raleigh, NC 27695, USA}
\affiliation{Joint Institute for Nuclear Astrophysics - Center for Evolution of the Elements, USA}
\email{iuroederer@ncsu.edu}
\author{Charli M. Sakari}
\affiliation{San Francisco State University, 1600 Holloway Avenue, San Francisco, CA, 94132 USA}
\email{sakaricm@sfsu.edu}

\linenumbers

\begin{abstract}

We present a detailed chemo-dynamical analysis of the chemically peculiar very metal-poor ([Fe/H]$=-2.22\pm 0.11$) Milky Way (MW) halo star, 2MASS J00101758$-$1735387 (hereafter \thestar), discovered by the $R$-Process Alliance. We derive abundances for 45 elements, including 28 neutron-capture elements, and confirm that \thestar\ is highly \rproc-enhanced, with [Eu/Fe]=$+1.30$. The light, $\alpha$, and Fe-peak element abundances ratios relative to Fe are all sub-solar, aside from C, Si, and most extremely, Ca. We investigate three possible formation scenarios based on the star's chemical abundances and orbital kinematics: whether it formed in the \textit{in-situ} MW halo, in a MW dwarf galaxy satellite, or in a satellite of the Magellanic Cloud system. Using the position of \thestar\ in the [Mg/Mn]--[Al/Fe] plane, the abundance pattern of both the light and heavy elements, and the kinematics, we conclude that \thestar\ most plausibly originated in a now-disrupted ultra-faint dwarf galaxy that was accreted into the MW halo. We examine potential chemical enrichment scenarios of \thestar, including one similar to the \rproc-enhanced UFD Reticulum II, or enrichment via a sub-Chandrasekhar Type Ia SNe, potentially supplemented by a Ca-rich gap transient and by a separate neutron-rich site that facilitated $r$-process enrichment. A nuclear cosmochronometric age of $12.5\pm4.2$ Gyr is derived from the [Th/Eu] ratio, consistent with an ancient origin in one of the earliest star-forming systems in the universe. 

\end{abstract}


\section{Introduction}\label{intro} 
 
The \textit{Gaia} \citep{GaiaCC22} era has revealed a wealth of stellar streams and substructures in and around the Milky Way (MW), confirming that our Galaxy's formation history has been shaped by the ongoing accretion of local dwarf galaxies and globular cluster systems \citep{chiba2000, carollo2007,bell2008,schlaufmann2012,helmi2017,roederer2018,Myeong2019,gudin2021,naidu2021,shank2023}. Signatures of the buildup of the halo from accreted systems are encoded in the chemical compositions of accreted metal-poor stars \citep{frebel-rev2018,roederer2018}. Detailed chemical and kinematic studies of these metal-poor stars residing in, e.g., the MW halo, could thus be the key to understanding our Galaxy's halo formation \citep[e.g.,][]{naidu2020}. 

Within this accreted halo framework, rapid neutron-capture process (\rproc) enhanced metal-poor stars serve as particularly powerful tracers, as the heavy elements ($Z>30$) preserved in their atmospheres can be linked to individual nucleosynthetic enrichment events in their progenitor systems \citep{frebel-rev2018}. Stars enhanced in \rproc\ elements have been suggested to form and originate in lower-mass ultra-faint dwarf (UFD) galaxies, which are then accreted into the MW halo \citep{roederer2018,gudin2021,naidu2022,shank2023}. This has been supported by evidence from both observations and simulations of \rproc-enhanced stars in UFDs \citep{ji2016Nat,roederer2016,brauer2019,hansen2020,Hirai2022,Hirai2025}. As such, these stars are particularly useful for understanding the primary astrophysical sites of the \rproc, which remain under discussion \citep[e.g.,][]{drout2017, shappee2017, cote2019, Kobayashi2020, Cowan2021}. 

In order to address outstanding questions about the \rproc\ and its astrophysical site, the $R$-Process Alliance\footnote{\href{https://sites.google.com/view/rprocessalliance/}{https://sites.google.com/view/rprocessalliance/}} (RPA) collaboration has been conducting one of the most extensive searches for \rproc-enriched stars in the MW. As a result, the RPA has assembled a sample of $\sim2,000$ metal-poor stars that contains a large number of \rproc-enhanced stars \citep[e.g.,][]{rpa1,rpa2,cain2018,gull2018,roederer2018a,placco2020,rpa3,rpa4,rpa5}. 

The initial analysis of the MW halo star 2MASS J00101758$-$1735387 (hereafter \thestar, also known as HE\,0007$-$1752) by the RPA indicated a chemical abundance pattern that could not be unambiguously attributed to either an \textit{in-situ} or \textit{ex-situ} origin \citep{rpa3,gull2021}. Here, we present a detailed chemical abundance and kinematic analysis of \thestar\ to further investigate its origin and birth environment. We consider three possible scenarios: formation in the $in$-$situ$ MW halo, accretion from a MW dwarf galaxy satellite, and an origin in a satellite of the Magellanic Cloud system. Determining which of these scenarios applies to \thestar\ can help answer questions about halo formation, such as how much of the MW's metal-poor halo was built through dwarf galaxy accretion \citep{frebel-rev2018, naidu2020}, as well as how the chemical abundances of stars that deviate from the typical halo patterns can test and highlight different nucleosynthetic pathways.

The \textit{in-situ} MW halo is composed of stars that formed within the MW's own potential well during the earliest phases of galaxy formation, before significant accretion events occurred \citep{helmi2018, belokurov2022}. At [Fe/H] $\lesssim-3$, the \textit{in-situ} halo population is dominated by ancient, metal-poor stars whose abundance patterns reflect enrichment by a small number of early core-collapse supernovae (CCSNe) \citep[e.g.,][]{frebel-rev2018}. \rproc-enhanced stars have been found throughout the in-situ halo \citep{rpa3, rpa4}, though their precise formation sites remain debated, and cases like \thestar\ offer a way to test this formation channel star-by-star.

\begin{deluxetable}{l r  r }
\tablewidth{0pt}
\tabletypesize{\scriptsize}
\tablecaption{\label{tab:stell_param} Fundamental, Kinematic, and Stellar Atmospheric 
Parameters of \thestar}

\tablehead{
\colhead{Parameter} & \colhead{Value}  & \nocolhead{NLTE}
}
\startdata
R.A. & 00:10:17.8 \\
Decl. & $-$17:35:37.3 \\
$l$ & 75.917$^{o}$\\
$b$ & $-$76.545$^{o}$\\
Parallax (mas) & 1.36 \\
\emph{V} (mag) & 11.539 \\
\emph{G} (mag) & 11.34 \\
\emph{K} (mag) & 9.64 \\
\emph{BP} (mag) & 11.70 \\
\emph{RP} (mag) & 11.80 \\
Observation Dates & Dec. 14 \& Dec. 15, 2020 \\
RV$_\mathrm{helio}^{N1}$ (\kms) & 190.1 $\pm 1.17$ \\
RV$_\mathrm{helio}^{N2}$ (\kms) & 190.2 $\pm 0.55$ \\
RV$_\mathrm{helio}^{Gaia}$ (\kms) & 191.8 $\pm 0.41$ \\
RV$_\mathrm{helio}^{RPA}$\tablenotemark{*} (\kms) & 191.8 \\
\hline
Stellar Parameters  & LTE\tablenotemark{\dag} & NLTE \\
\hline
\teff\ (K)   & 5430     $\pm$ 100  & 5037    $\pm$ 78  \\
\logg\ (cgs) & 3.05     $\pm$ 0.20 & 2.54    $\pm$ 0.22 \\
\feh\        & $-$2.22  $\pm$ 0.11 & $-$2.52 $\pm$ 0.06 \\
\vt\ (\kms)  & 1.33     $\pm$ 0.20 & 1.42    $\pm$ 0.08 \\
\enddata
\tablenotemark{*}{RPA parameters from \citet{rpa3}} 
\tablenotetext{\dag}{Final adopted parameters}
\end{deluxetable}




\begin{figure*}
\begin{center}
\includegraphics[scale=0.42]{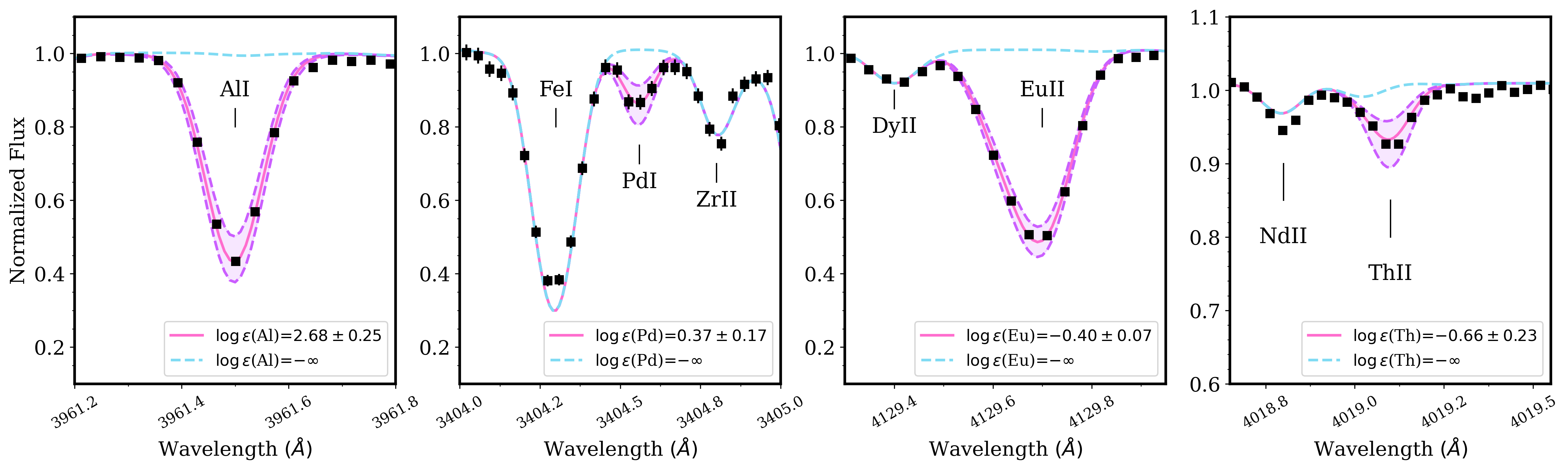}
\caption{\label{fig:spectrum} Observed spectrum (black dots) and representative absorption line fits (pink lines) for \thestar\. The blue dashed lines simulate the spectrum without any contribution of the respective element to the fit ($\log\epsilon$(X)$=-\infty$), and the purple region shows the uncertainty around the fits.}
\end{center}
\end{figure*}  

An alternative \textit{ex-situ} origin could be accretion from a satellite galaxy, either a classical MW dwarf spheroidal or a UFD galaxy, both of which are known to host \rproc-enhanced metal-poor stars \citep{ji2016Nat, roederer2016, hansen2020}. Stars accreted from these systems are expected to show distinct chemical signatures relative to those that formed in the \textit{in-situ} halo, including lower [Al/Fe], sub-solar [Mg/Mn], and systematically under-enhanced Fe-peak abundances reflecting the lower star-formation efficiency and truncated enrichment histories of their progenitor systems \citep{hawkins2015, das2020, Simon2019rev}. More recently, \rproc-enhanced stars have been identified in the Magellanic Cloud system  \citep{reggiani2021}, and several UFDs have been confirmed as LMC satellites accreted into the MW alongside the Magellanic system \citep{kallivayalil2018,ji2020car}. Stars originating in LMC satellites exhibit kinematic associations with the Magellanic cloud systems \citep[i.e.,][]{chandra2023}.

The remainder of this paper is outlined as follows.  In Section \ref{sec:obs}, we describe the observations taken of \thestar. 
In Sections \ref{sec:stell_param} and \ref{abundanalysis} we present the methods and results employed to obtain both the stellar parameters and abundances. We review the chemical abundances and kinematic properties of this star in Sections \ref{sec:abund} and \ref{sec:kinematics}. We discuss potential chemical enrichment scenarios in Section \ref{sec:chemenrichment}. We then compare the potential birth environments of \thestar\ in Sections \ref{sec:mwstars} and  \ref{sec:mwsatellite}. In Section \ref{sec:age}, we use nuclear cosmo-chronometry to obtain an age estimate for \thestar. Finally, we summarize our findings in Section \ref{sec:conc}.

\section{Observations, Data Reduction, and Radial Velocity Determination}\label{sec:obs}

\thestar\ was identified and vetted as a metal-poor candidate by the RPA collaboration \citep{placco2018} from the Large Sky Area Multi-Object Fibre Spectroscopic Telescope (LAMOST) survey \citep{LAMOST}. A snapshot spectrum of the star was observed with the Magellan Inamori Kyocera Echelle (MIKE) spectrograph \citep{bernstein2003} on the $Magellan$-Clay Telescope at Las Campanas Observatory on October 14, 2016, for 1200\,s with the 0\farcs7 slit and 2$\times$2 binning. The analysis of this spectrum was presented in \cite{rpa3}, confirming the star to be \rproc\ enhanced with $\mathrm{[Eu/Fe]} = +1.35$ and $\mathrm{[Ba/Eu]} = -0.92$. 

Consequently, a portrait spectrum was obtained with the Magellan/MIKE spectrograph on December 14 and 15, 2020, with the 0\farcs5 slit and 2$\times$1 binning, yielding a resolving power of $R\sim$ 60,000 and 50,000 in the blue and red wavelength regimes, respectively. The total exposure time was 4.0 hours, separated into 13 exposures (8 $\times$ 1200\,s on Dec 14, and 5 $\times$ 1200\,s on Dec 15). All data were reduced with the CarPy MIKE pipeline \citep{kelson2000,kelson2003}, and heliocentric radial velocities were determined for each night by cross-correlating the spectra against a spectrum of the benchmark star HD 122563 \citep{gaia2018}, yielding 190.1 \kms\ on Night 1, and 190.2 \kms\ on Night 2. These radial velocities are in good agreement with the radial velocities found in \citet{Gaiasummary2022} and \citet{rpa3}.

The final spectrum has signal-to-noise ratios (SNR) of $\sim50$ at 3500\,{\AA} and $\sim190$ at 4000\,{\AA}, with select lines shown in Figure~\ref{fig:spectrum}. The properties of \thestar\ are listed in Table~\ref{tab:stell_param}.

\section{Stellar Atmospheric Parameters}\label{sec:stell_param}

Stellar parameters were determined following the same procedures as adopted for other RPA-discovered stars. For further details, see \citet{roederer2022,shah2024}. Here we briefly summarize the main steps undertaken. The effective temperature (\teff) of the star was determined photometrically from the $Gaia$ $G$, $BP$, $RP$, and the $K$ magnitudes using the color-temperature relations from \citet{mucciarelli2021}. The $Gaia$ magnitudes were de-reddened using $E(B-V) = 0.0295$ obtained from the \cite{Schlafly2011} dust maps, with a correction from \cite{Bonifacio2000}. To de-redden the $K$ magnitude, we use the extinction coefficient from \cite{mccall2004}. Following the \teff\ determination, we calculate \logg\ from the fundamental relation:
\vspace{1em}

$\log g = 4 \log $ \teff $ + \log$(M/M$_{\odot}$) $-$ 10.61 + 0.4$\cdot(BC_{V} \nonumber\\ + m_{V} - 5\log (d) + 5 - 3.1 \cdot E(B-V) - M_{\rm bol,\odot}).$ 
\vspace{1em}

\begin{deluxetable}{l c c r r r c r r c c}
\tabletypesize{\scriptsize}
\tablecaption{\label{tab:chemabun} Chemical Abundances of \thestar}

\tablehead{
\\
\colhead{El.} & \colhead{N} & \colhead{$\log\epsilon$(X)$_{\odot}$} & \colhead{$\log\epsilon$(X)} & \colhead{[X/H]} & \colhead{[X/Fe]\tablenotemark{\dag}} & \colhead{$\sigma_{rand}$} & \colhead{$\log\epsilon$(X)} & \colhead{[X/Fe]} & \colhead{$\sigma_{rand}$} & \colhead{$\sigma_{tot}$} 
\\
\colhead{} & \colhead{} & \colhead{} & \colhead{LTE} & \colhead{LTE} & \colhead{LTE} & \colhead{LTE [dex]} & \colhead{NLTE} & \colhead{NLTE} & \colhead{NLTE [dex]} & \colhead{[dex]} }
\startdata
  CH         &  1    &  8.43   &   6.27  &  $-$2.16 &    0.06   & 0.07   &   \nodata\ &    \nodata\  &  \nodata\ & 0.15 \\
  NH         &  1    &  7.83   &   5.33  &  $-$2.50 &   $-$0.27   &  \nodata\   &   \nodata\ &    \nodata\  &  \nodata\ & 0.18 \\
\ion{Na}{i}  &  2    &  6.24   &   3.37  &  $-$2.87 &   $-$0.64   & 0.06   &  3.07 &  $-$0.96  & 0.02 &  0.12 \\
\ion{Mg}{i}  &  7    &   7.60  &   4.99  & $-$2.61   &  $-$0.38    & 0.07 &  4.98   &   $-$0.40  &   0.07 & 0.06 \\
\ion{Al}{i}  &  2    &  6.45   &   2.68  &  $-$3.77 &   $-$1.55   &  0.01   &   3.13  &   $-$1.09   &  \nodata\ & 0.25 \\
\ion{Si}{i}  &  2    &  7.51   &   5.37  &  $-$2.14 &    0.08   &  \nodata\   &   5.29  &    0.01   &  0.32 & 0.05 \\
\ion{K}{i}   &  1    &  5.03   &   2.42  &  $-$2.61 &   $-$0.39   &  \nodata\  &   \nodata\ &    \nodata\  &  \nodata\ & 0.21 \\
\ion{Ca}{i}  & 25    &  6.34   &   4.29  &  $-$2.05 &    0.17   & 0.05   &  4.27 &   0.15  & 0.05 & 0.05 \\
\ion{Sc}{ii} &  8    &  3.15   &   0.06  &  $-$3.09 &   $-$0.87   & 0.04   &   \nodata\ &    \nodata\  &  \nodata\ & 0.06 \\
\tii\        & 18    &  4.95   &   2.53  &  $-$2.42 &   $-$0.19   & 0.08   &  2.78   &   0.05  &   0.16 & 0.09 \\
\tiii\       & 29    &  4.95   &   2.55  &  $-$2.40 &   $-$0.18   & 0.07   &  2.64 &  $-$0.09  &  0.10 &  0.06 \\
\ion{V}{i}   &  1    &  3.93   &   0.80  &  $-$3.13 &   $-$0.90   &  \nodata\   &   \nodata\ &    \nodata\  &  \nodata\ & 0.44 \\
\ion{V}{ii}  &  4    &  3.93   &   1.01  &  $-$2.92 &   $-$0.70   & 0.02   &   \nodata\ &    \nodata\  &  \nodata\ &  0.08 \\
\ion{Cr}{i}  & 10    &  5.64   &   3.23  &  $-$2.41 &   $-$0.19   & 0.02   &   \nodata\ &    \nodata\  &  \nodata\ & 0.07 \\
\ion{Cr}{ii} &  4    &  5.64   &   3.35  &  $-$2.29 &   $-$0.06   & 0.08   &   \nodata\ &    \nodata\  &  \nodata\ & 0.05 \\
\ion{Mn}{i}  &  8    &  5.43   &   2.44  &  $-$2.99  &  $-$0.77    & 0.04   &  2.72   &  $-$0.49 &   0.07 & 0.08 \\
\ion{Mn}{ii} &  3    &  5.43   &   2.57  &  $-$2.86 &   $-$0.64   & 0.03   &   \nodata\ &    \nodata\  &  \nodata\ & 0.07 \\
\fei\        &210    &  7.50   &   5.28  &  $-$2.22 &    0.00    & 0.11   &   \nodata\ &    \nodata\  &  \nodata\ &   0.11 \\
\feii\       & 16    &  7.50   &   5.28  &  $-$2.22 &    0.01   & 0.08   &   \nodata\ &    \nodata\  &  \nodata\ &   0.08 \\
\ion{Co}{i}  & 20    &  4.99   &   2.56  &  $-$2.43 &   $-$0.20    &  0.10   &  3.15   &   0.38  &   0.15 & 0.09 \\
\ion{Ni}{i}  & 26    &  6.22   &   3.52  &  $-$2.70  &  $-$0.48    & 0.13   &  3.72   &  $-$0.28 &   0.22 & 0.07 \\
\ion{Zn}{i}  &  1    &  4.56   &   1.88  &  $-$2.68 &   $-$0.46   &  \nodata\   &   \nodata\ &    \nodata\  &  \nodata\ & 0.20 \\
\ion{Sr}{ii} &  3    &  2.87   &   0.91  &  $-$1.96 &    0.26   & 0.16   &  0.98   &   0.33  &   0.16 & 0.13 \\
\ion{Y}{ii}  & 17    &  2.21   &   0.13  &  $-$2.08 &    0.14   & 0.11   &  0.07 &   0.08  & 0.14 & 0.05 \\
\ion{Zr}{ii} & 17    &  2.58   &   0.81  &  $-$1.77 &    0.45   & 0.08   &   \nodata\ &    \nodata\  &  \nodata\ & 0.07 \\
\ion{Mo}{i}  &  1    &  1.88   &   0.32   & $-$1.56  &   0.66    &  \nodata\   &   \nodata\ &    \nodata\  &  \nodata\ & 0.23 \\
\ion{Ru}{i}  &  1    &  1.75   &   0.63  &  $-$1.12 &    1.10   &  \nodata\   &   \nodata\ &    \nodata\  &  \nodata\ & 0.25 \\
\ion{Rh}{i}  &  1    &  0.91   &  $-$0.20  &  $-$1.11 &    1.12   &  \nodata\   &   \nodata\ &    \nodata\  &  \nodata\ & 0.27 \\
\ion{Pd}{i}  &  2    &  1.57   &   0.37  &  $-$1.20  &   1.03    & 0.04   &   \nodata\ &    \nodata\  &  \nodata\ & 0.17 \\
\ion{Ag}{i}  &  1    &  0.94   &  $-$0.60  &  $-$1.54 &    0.68   &  \nodata\   &   \nodata\ &    \nodata\  &  \nodata\ & 0.23 \\
\ion{Ba}{ii} &  5    &  2.18   &   0.80  &  $-$1.38 &    0.85   & 0.11   &  0.65   &   0.69   &  0.26 & 0.14 \\
\ion{La}{ii} & 21    &  1.10   &  $-$0.17  &  $-$1.27 &    0.95   & 0.06   &   \nodata\ &    \nodata\  &  \nodata\ & 0.05 \\
\ion{Ce}{ii} & 32    &  1.58   &   0.14  &  $-$1.44 &    0.78   & 0.08   &   \nodata\ &    \nodata\  &  \nodata\ &  0.07 \\
\ion{Pr}{ii} &  8    &  0.72   &  $-$0.37  &  $-$1.09  &   1.13    & 0.14   &   \nodata\ &    \nodata\  &  \nodata\ & 0.07 \\
\ion{Nd}{ii} & 58    &  1.42   &   0.24  &  $-$1.18 &    1.04   & 0.08   &  0.33 &   1.13  &  0.10 & 0.06 \\
\ion{Sm}{ii} & 56    &  0.96   &  $-$0.07  &  $-$1.03 &    1.19    & 0.07   &   \nodata\ &    \nodata\  &  \nodata\ & 0.07 \\
\ion{Eu}{ii} &  6    &  0.52   &  $-$0.40  &  $-$0.92 &    1.30   & 0.12   & $-$0.28 &   1.42  &  0.30 & 0.07 \\
\ion{Gd}{ii} & 19    &  1.07   &   0.12  &  $-$0.95 &    1.27   & 0.07   &   \nodata\ &    \nodata\  &  \nodata\ & 0.07 \\
\ion{Tb}{ii} &  2    &  0.30   &  $-$0.62   & $-$0.92  &   1.30    & 0.08   &   \nodata\ &    \nodata\  &  \nodata\ & 0.18 \\
\ion{Dy}{ii} & 24    &  1.10   &   0.39   & $-$0.71  &   1.51    & 0.13   &   \nodata\ &    \nodata\  &  \nodata\ & 0.06 \\
\ion{Ho}{ii} &  6    &  0.48   &  $-$0.41  &  $-$0.89 &    1.34   & 0.05   &   \nodata\ &    \nodata\  &  \nodata\ & 0.08 \\
\ion{Er}{ii} & 10    &  0.92   &   0.10  &  $-$0.82  &   1.41    & 0.09   &   \nodata\ &    \nodata\  &  \nodata\ & 0.06 \\
\ion{Tm}{ii} &  4    &  0.10   &  $-$0.68  &  $-$0.78 &    1.44   & 0.08   &   \nodata\ &    \nodata\  &  \nodata\ & 0.07 \\
\ion{Yb}{ii} &  1    &  0.84   &   0.03  &  $-$0.81 &    1.41   & \nodata\   &   \nodata\ &    \nodata\  &  \nodata\ & 0.24 \\
\ion{Lu}{ii} &  1    &  0.10   &  $-$0.72   & $-$0.82  &   1.40    & \nodata\   &   \nodata\ &    \nodata\  &  \nodata\ & 0.24 \\
\ion{Hf}{ii} &  4    &  0.85   &  $-$0.30  &  $-$1.15 &    1.08   &  0.10   &   \nodata\ &    \nodata\  &  \nodata\ & 0.07 \\
\ion{Os}{i}  &  1    &  1.40   &   0.63  &  $-$0.77 &    1.45    &  \nodata\   &   \nodata\ &    \nodata\  &  \nodata\ & 0.23 \\
\ion{Ir}{i}  &  1    &  1.38   &   0.56  &  $-$0.82  &   1.40    &  \nodata\   &   \nodata\ &    \nodata\  &  \nodata\ & 0.26 \\
\ion{Pb}{i}  &  1    &  1.75 & $\leq$0.49 & $-$1.26  &  0.96    &  \nodata\   &   \nodata\ &    \nodata\  &  \nodata\ & \nodata\\
\ion{Th}{ii} &  1    &  0.02   & $-$0.66 &  $-$0.68  &  1.54    &  \nodata\   &   \nodata\ &    \nodata\  &  \nodata\ & 0.23 \\
\enddata
\tablecomments{Solar abundances (Column 3) are adopted from \citet{asplund2009}. The references for NLTE model atoms and departure coefficient grids are: Na \citep{na_al_nlte}, Mg \citep{mg_nlte}, Al \citep{na_al_nlte}, Si \citep{si_nlte}, Ca \citep{mashonkina2017}, Ti \citep{ti_nlte}, Mn \citep{mn_nlte}, Co \citep{co_nlte}, Ni \citep{o_ni_nlte}, Sr \citep{sr_nlte}, Y \citep{y_nlte, y_eu_nlte}, Ba \citep{ba_nlte}, Nd \citep{dixon2025}, and Eu \citep{y_eu_nlte}.}
\tablenotetext{\dag}{We adopt the LTE abundances for the final analysis yet report NLTE abundances where available.}
\end{deluxetable}

Here, $M$ is the mass of the star, assumed to be 0.8 $\pm$ 0.08 M$_{\odot}$.  BC$_V$ is the bolometric correction in the $V$ band \citep{casagrande2014}, $m_V$ is the apparent $V$magnitude, $d$ is the distance in parsecs from \citet{bailorjones2021}, and M$_{\rm bol,\odot}$ is the solar bolometric magnitude, 4.75.  The constant 10.61 is calculated from the solar constants log(\teff)$_{\odot}$ = 3.7617 and \logg$_{\odot}$ = 4.438. Finally, the model atmosphere metallicity $\mathrm{[M/H]}$ and microturbulence \vmic\ are determined from equivalent width (EW) measurements of \ion{Fe}{i} and \ion{Fe}{ii} lines, using the $\mathrm{[FeII/H]}$ abundance as model atmosphere metallicity. The final stellar parameters are listed in Table \ref{tab:stell_param}, with \teff= 5430 K, \logg=3.05, \feh=$-2.22$, and \vmic=1.33 \kms. Following \citet{placco2017}, we adopt uncertainties of 100 K, 0.2 cgs, and 0.2 \kms\ for our stellar parameters of \teff, \logg, and \vmic, respectively. 


For warm (\teff $>5000$\,K) metal-poor giant stars, discrepancies of parameters derived using Fe excitation and ionization balance as compared to parameters derived using fundamental equations have been attributed to non-LTE (NLTE) effects \citep{lind2012, bergemann2014, Ezzeddine2017}. We therefore also derived stellar parameters using \texttt{LOTUS} \citep{lotus}, a 1D NLTE tool that determines \teff, \logg, \feh, and \vt\ from EW measurements of \fei\ and \feii\ lines via a generalized curve-of-growth method and global differential evolution optimization with MCMC-based uncertainties. The NLTE stellar parameters for \thestar\ are \teff\ = $5036 \pm 78$\,K, \logg\ = $2.54 \pm 0.22$, \feh\ = $-2.52 \pm 0.06$, and \vt\ = $1.42\pm0.08$\,\kms. The derived stellar parameters are summarized in Table~\ref{tab:stell_param}.

\section{Abundance Analysis}\label{abundanalysis}

\begin{deluxetable}{l r r r r}
\tablewidth{12pt}
\tabletypesize{\scriptsize}
\tablecaption{\label{tab:errors} Systematic Abundance Uncertainties for \thestar}
\tablehead{
\\
\colhead{}& \colhead{[X/Fe]} & \colhead{$\Delta$ \teff} & \colhead{$\Delta$ \logg} & \colhead{$\Delta$ \vt}
\\
\colhead{}  & \colhead{}  & \colhead{$\pm$100\,K} & \colhead{$\pm$0.2\,cgs} & \colhead{$\pm$0.2\,km/s}
\\
\colhead{}  & \colhead{}  & \colhead{[dex]} & \colhead{[dex]} & \colhead{[dex]}}
\startdata
  CH & $+0.14$   &  $+0.09$  &  $-0.07$   &  $+0.04$ \\
  NH & $-0.24$   &  $+0.10$  &  $-0.08$   &  $+0.04$ \\
\ion{Na}{i} & $-0.64$   &  $+0.01$  &  $  0.00$   &  $-0.03$ \\
\ion{Mg}{i} & $-0.38$   &  $-0.03$  &  $-0.02$   &  $+0.02$ \\
\ion{Al}{i} & $-1.56$   &  $+0.01$  &  $-0.01$   &  $-0.02$ \\
\ion{Si}{i} & $+0.08$   &  $-0.06$  &  $+0.03$   &  $+0.04$ \\
\ion{K}{i}  & $-0.39$   &  $-0.02$  &  $+0.01$   &  $+0.03$ \\
\ion{Ca}{i} & $+0.17$   &  $-0.03$  &  $  0.00$   &  $+0.02$ \\
\ion{Sc}{ii} & $-0.86$   &  $-0.03$  &  $+0.06$   &  $+0.04$ \\
\tii & $-0.19$   &  $+0.01$  &  $  0.00$   &  $+0.03$ \\
\tiii & $-0.17$   &  $-0.05$  &  $+0.06$   &  $-0.01$ \\
\ion{V}{i}  & $-0.85$   &  $  0.00$  &  $+0.02$   &  $+0.04$ \\
\ion{V}{ii} & $-0.69$   &  $-0.05$  &  $+0.07$   &  $+0.05$ \\
\ion{Cr}{i} & $-0.19$   &  $+0.01$  &  $  0.00$   &  $+0.01$ \\
\ion{Cr}{ii} & $-0.06$   &  $-0.09$  &  $+0.08$   &  $+0.03$ \\
\ion{Mn}{i} & $-0.77$   &  $  0.00$  &  $  0.00$   &  $  0.00$ \\
\ion{Mn}{ii} & $-0.63$   &  $-0.07$  &  $+0.07$   &  $-0.01$ \\
\fei & $  0.00$   &  $  0.00$  &  $  0.00$   &  $  0.00$ \\
\feii & $+0.02$   &  $-0.08$  &  $+0.07$   &  $  0.00$ \\
\ion{Co}{i} & $ -0.20$   &  $+0.04$  &  $-0.01$   &  $+0.01$ \\
\ion{Ni}{i} & $-0.48$   &  $+0.02$  &  $-0.01$   &  $-0.03$ \\
\ion{Zn}{i} & $-0.46$   &  $-0.04$  &  $+0.04$   &  $+0.04$ \\
\ion{Sr}{ii} & $+0.27$   &  $-0.01$  &  $+0.04$   &  $-0.02$ \\
\ion{Y}{ii} & $+0.15$   &  $-0.04$  &  $+0.07$   &  $  0.00$ \\
\ion{Zr}{ii} & $+0.47$   &  $-0.03$  &  $+0.06$   &  $+0.03$ \\
\ion{Mo}{i} & $+0.64$   &  $-0.03$  &  $+0.08$   &  $+0.05$ \\
\ion{Ru}{i} & $ 1.09$   &  $+0.03$  &  $-0.01$   &  $+0.04$ \\
\ion{Rh}{i} & $ 1.05$   &  $+0.03$  &  $+0.03$   &  $+0.11$ \\
\ion{Pd}{i} & $ 1.03$   &  $+0.04$  &  $-0.01$   &  $+0.03$ \\
\ion{Ag}{i} & $+0.67$   &  $+0.01$  &  $-0.03$   &  $+0.04$ \\
\ion{Ba}{ii} & $+0.85$   &  $  0.00$  &  $+0.04$   &  $-0.09$ \\
\ion{La}{ii} & $+0.96$   &  $-0.03$  &  $+0.06$   &  $+0.03$ \\
\ion{Ce}{ii} & $+0.79$   &  $-0.03$  &  $+0.06$   &  $+0.03$ \\
\ion{Pr}{ii} & $ 1.12$   &  $-0.03$  &  $+0.07$   &  $+0.04$ \\
\ion{Nd}{ii} & $ 1.06$   &  $-0.03$  &  $+0.06$   &  $+0.03$ \\
\ion{Sm}{ii} & $  1.20$   &  $-0.03$  &  $+0.06$   &  $+0.03$ \\
\ion{Eu}{ii} & $ 1.31$   &  $-0.04$  &  $+0.07$   &  $+0.04$ \\
\ion{Gd}{ii} & $ 1.29$   &  $-0.03$  &  $+0.06$   &  $+0.04$ \\
\ion{Tb}{ii} & $ 1.33$   &  $-0.04$  &  $+0.07$   &  $+0.05$ \\
\ion{Dy}{ii} & $ 1.52$   &  $-0.03$  &  $+0.06$   &  $  0.00$ \\
\ion{Ho}{ii} & $ 1.35$   &  $-0.03$  &  $+0.05$   &  $+0.04$ \\
\ion{Er}{ii} & $ 1.42$   &  $-0.04$  &  $+0.07$   &  $  0.00$ \\
\ion{Tm}{ii} & $ 1.45$   &  $-0.02$  &  $+0.08$   &  $+0.05$ \\
\ion{Yb}{ii} & $ 1.45$   &  $-0.04$  &  $+0.07$   &  $  0.00$ \\
\ion{Lu}{ii} & $ 1.42$   &  $-0.07$  &  $+0.07$   &  $+0.04$ \\
\ion{Hf}{ii} & $ 1.08$   &  $-0.04$  &  $+0.06$   &  $+0.04$ \\
\ion{Os}{i} & $  1.50$   &  $  0.00$  &  $-0.07$   &  $+0.02$ \\
\ion{Ir}{i} & $ 1.42$   &  $+0.05$  &  $-0.02$   &  $+0.01$ \\
\ion{Th}{ii} & $ 1.43$   &  $-0.03$  &  $+0.06$   &  $+0.02$ \\
\enddata
\end{deluxetable} 

Abundances were derived by fitting synthetic spectra to each individual observed spectral line via a $\chi^2$ minimization optimizing the abundance of that line, the local continuum, and a Gaussian smoothing kernel using \texttt{SMHr} \citep{casey2014,smhr}. \texttt{SMHr} incorporates 1D LTE ATLAS-Kurucz model atmospheres \citep{castelli2004} and the LTE radiative transfer code \texttt{MOOG}\footnote{\url{https://github.com/alexji/moog17scat}} \citep{sneden1973} with Rayleigh scattering \citep{sobeck2011}. The atomic line data (wavelengths, excitation potentials, and oscillator strengths) were adopted from the line list generated using \texttt{linemake} \citep{linemake} in \citet{roederer2018a} and references therein. We report the line information in Table~\ref{tab:linemeasurements}, including the wavelengths, species, excitation potentials, oscillator strengths, equivalent widths (where available), and absolute LTE abundances. All abundance ratios [X/H] are reported relative to the Solar abundances of \citet{asplund2009}. The chemical abundances of \thestar\ in LTE are listed in Table~\ref{tab:chemabun}.


We also derived NLTE abundances for 14 elements for which model atoms and grids of NLTE departure coefficients are available. We used the spectral synthesis code \texttt{Turbospectrum} with NLTE capabilities \citep{alvarez1998, plez2012, gerber2023, plez2025}, and its Python wrapper \texttt{TSFitPy} \citep{y_nlte}, to generate synthetic spectra and determine the best-fit abundances. The main inputs to the NLTE spectral synthesis are model atmospheres, line lists, model atoms, and pre-calculated grids of NLTE departure coefficients. \texttt{TSFitPy} uses the 1D MARCS grid of stellar model atmospheres \citep{gustafsson1975, gustafsson2008}. The line lists are taken from the Gaia-ESO line list \citep{heiter2021} for wavelengths between 4,750~\AA\ and 6,850~\AA, and from the VALD\footnote{\url{https://vald.astro.uu.se/}} database for wavelengths shorter than 4,750~\AA\ or longer than 6,850~\AA. 
The pre-computed grids of departure coefficients describe the departures of the atomic level populations from their LTE values.
The references for model atoms and departure coefficients are listed at the bottom of Table~\ref{tab:chemabun}. 
We report the NLTE abundances and the standard deviations among lines in Table~\ref{tab:chemabun} as well. Besides the line-by-line scatter, the typical $1\sigma$ fitting uncertainty for a single line is $\sim 0.05$~dex, largely independent of the elements in question. 

Systematic uncertainties of the chemical abundances as a function of stellar parameters were estimated following \citet{placco2017}, by varying each atmospheric parameter independently by $+100$ K for \teff, $+0.2$ dex for \logg, and $+0.2$ \kms\ for \vt. The subsequent changes in each elemental abundance are listed in Table~\ref{tab:errors}. The total uncertainties were estimated using the error-propagation method described in \citet{ji2020s5} with updates described in \citet{atzberger2025}. To summarize, we found the stellar parameter uncertainties by assigning each line a statistical error $e_i$ and a species-wide systematic floor, $s_x$, combined to create a total uncertainty of $\sigma_i^2 = e_i^2 + s_x^2$. This term is combined with the shared uncertainty dependence on the stellar parameters (\teff, \logg, \vt, and \feh). For a given species, we construct an $4\times4$ covariance matrix, $\Sigma=$diag($\sigma_i^2) + \delta\rho\delta^T$. The off-diagonal terms account for correlated line behavior using the parameter correlation matrix ($\rho$) paired with the lines' measured sensitivities ($\delta$). Inverting $\Sigma$ yields effective per-line weights used to calculate mean abundances and allows us to propagate correlated errors into final [X/Fe] ratios without overestimation. The total uncertainties are reported in Table~\ref{tab:chemabun} as $\sigma_{tot}$. 



\section{Results}
\subsection{Chemical Abundances}\label{sec:abund}


We determined abundances and upper limits of \thestar\ for CH, NH, the light elements (Na, Al, K), $\alpha$-elements (Mg, Si, Ca), Fe-peak elements (Sc, Ti, V, Cr, Mn, Fe, Co, Ni, Zn), and 28 neutron-capture elements (Z $ >30$); upper limits are reported where detections were not possible. The LTE abundances, NLTE abundances, and upper limits are listed in Table~\ref{tab:chemabun} along with the standard deviation among lines ($\sigma_{rand}$) for each element. 

We compare our final abundances to those derived by \citet{gull2021}. Only one abundance, Al, deviates by more than $\pm$0.2 dex: we derive [Al/Fe]$=-1.55$ from two lines, compared to [Al/Fe]$=-0.58$ with one line from \citet{gull2021}. We detect \ali\ from the lines at 3944 and 3961\,\AA, which yield consistent abundances. Both lines lie close to the \ion{Ca}{II} H and K lines, where continuum placement can strongly affect the derived abundance \citep{Timmes1995}. However, as the two lines were fit independently and yield consistent results within 0.01 dex, we show a reliable fit for 3961\,{\AA} in Figure \ref{fig:spectrum}.

Figure~\ref{fig:abundplot} shows the elemental abundances of \thestar\ along with four other groups of systems: MW halo stars (blue circles), dSphs (pink downward triangles), UFD satellites of the MW (purple triangles), and UFD satellites of the LMC (open circles). Other ``Iron-rich Metal-poor'' stars are shown where available in green stars \citep{venn2012,sakari2019,Reggiani2023} based on the definition put forth by \citet{Reggiani2023}. Individual references are given in the figure caption.


\subsubsection{Carbon and Light Elements: C, O, Na, Al and K}\label{sec:light-el_abund}

The carbon abundance was determined via spectrum synthesis from the molecular CH $G$-band at $4220-4323$\,{\AA} using the linelist of \citet{masseron2014}, yielding an abundance of [C/Fe] $=0.06$. The nitrogen abundance was found from the NH feature at 3360\,{\AA}, yielding [N/Fe] = $-0.27$.


Oxygen is notoriously difficult to detect in metal-poor stars and both the forbidden [\ion{O}{I}] lines at 6300 and 6363\,\AA\ were not detectable in \thestar.

Sodium was derived from the \nai\ doublet at 5889 and 5895\,\AA, yielding [Na/Fe] $ =-0.64$. Aluminum was detected from two lines at 3961 and 3944\,\AA, yielding [Al/Fe] $ =-1.55$. Potassium was detected from a single line at 7699\,\AA, yielding [K/Fe] $ =-0.39$.

We found an NLTE sodium abundance of [Na/Fe]$_{\text{NLTE}} $=$-0.96$ and an NLTE aluminum abundance of [Al/Fe]$_{\text{NLTE}}= -1.09 $. \thestar\ is systematically sub-solar and deficient in [X/Fe] ratios compared to other halo stars for the light , except for C. 

\subsubsection{The $\alpha$-elements}\label{sec:alpha-el_abund}

$\alpha$-element abundances were derived for \mgi, \sii, and \cai. A Mg abundance of [Mg/Fe] = $-0.38$ was determined using 7 \mgi\ lines. Two Si lines were detected at 3905\,{\AA} and 4102\,{\AA}, with a derived abundance of [Si/Fe] = +0.08. 

We derive a super-solar Ca abundance of [Ca/Fe] = $+0.17$ using 23 lines. In Section \ref{sec:chemenrichment}, we discuss the possible origins of this mild Ca enhancement. Aside from Ca, the $\alpha$-element abundances of \thestar\ are systematically lower than those of other MW halo stars.

The NLTE abundances derived for the $\alpha$-elements are [Mg/Fe]$_{\text{NLTE}} $ = $-0.40$, [Si/Fe]$_{\text{NLTE}} $ = $0.01$, and [Ca/Fe]$_{\text{NLTE}}$ = $+0.15$. The differences between the NLTE and LTE abundances for the $\alpha$-elements are less than 0.1 dex and do not alter the conclusions drawn for \thestar.

\subsubsection{Iron-peak Elements}\label{sec:fepeak_abund}

\begin{figure*}[ht!]
\begin{center}
\vspace*{-0cm}
\hspace*{-1cm}
\includegraphics[scale=0.575]{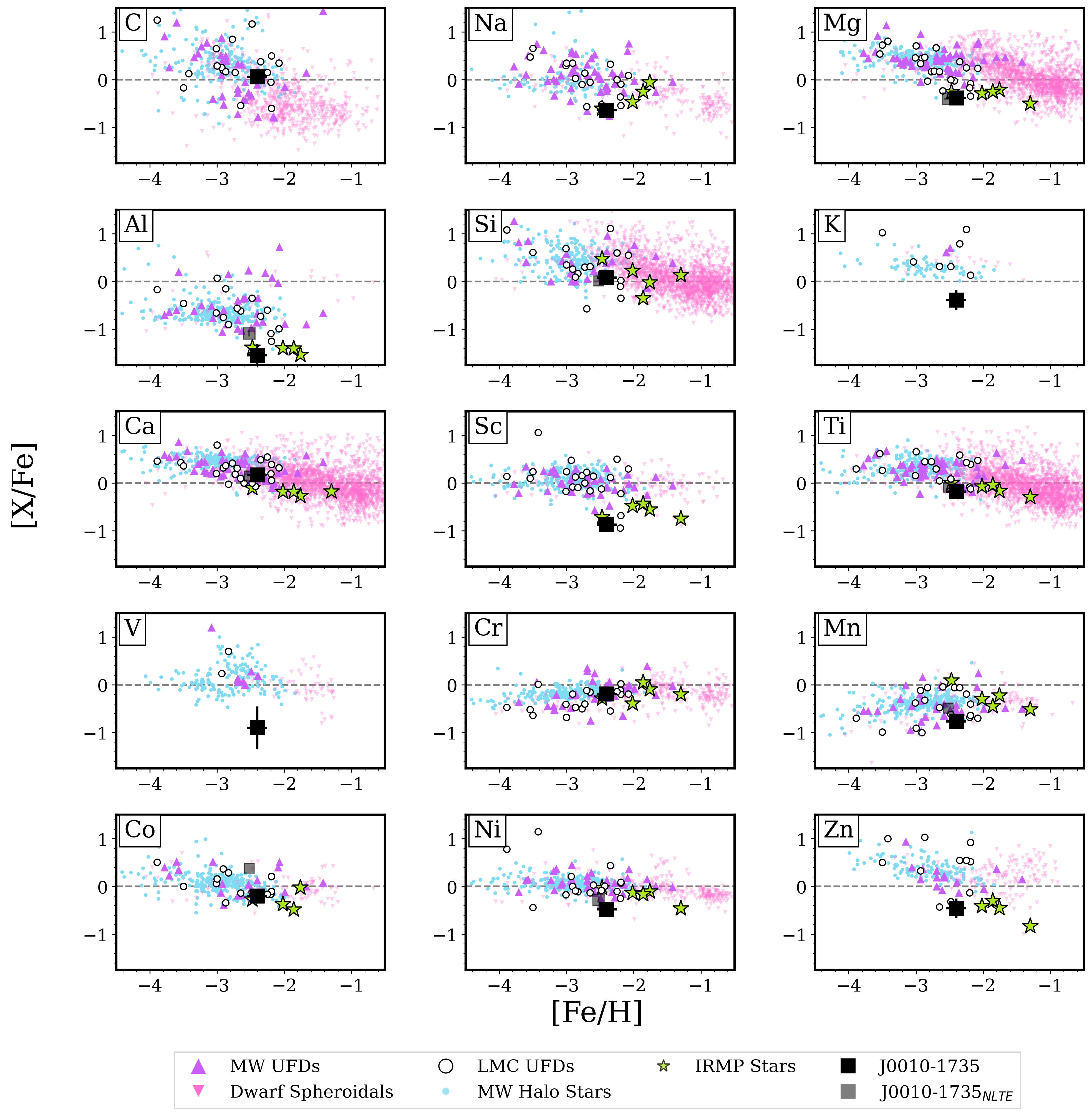}
\caption{\label{fig:abundplot} [X/Fe] ratios of the light, $\alpha$, and Fe-peak elements, as a function of \feh, for \thestar\ (black square), MW halo stars (blue  circles), various dSphs (blue upside-down triangles), MW UFDs (purple triangles) and LMC UFDs (open circles). Consistently, \thestar\ shows lower abundances with respect to both MW Halo and UFD stars.  The references for the abundances for the MW UFDs are: Hercules \citep{francois2016,Koch2008}, Bootes I \citep{frebel2016boo,norris2010,feltzing2009,ishigaki2014,gilmore2013}, Bootes II \citep{francois2016,ji2016bootes}, Ursa Major II \citep{frebel2010uma}, Coma Berenices \citep{frebel2010uma}, Leo IV \citep{simon2010leo,francois2016}, Canes Venatici I \& Canes Venatici II \citep{francois2016}, Tucana II \citep{ji2016tuc}, Tucana III \citep{marshall2019} and Grus II \citep{hansen2020}. The dSph references are Segue I \citep{frebel2014}, Segue II \citep{roederer2014seg}, Sculptor \citep{shetrone2003,geisler2005}, Ursa Minor \citep{shetrone2003,cohen2010}, Draco \citep{shetrone2003,cohen2010}, and Fornax \citep{shetrone2003,cohen2010}. The LMC UFD references are Reticulum II \citep{ji2016b,roederer2016}, , Horologium I \citep{nagasawa2018}, Carina II \& Carina III \citep{ji2020car}. The references for the MW stars are \citet{yong2013} and \citet{roederer2014}; the references for the IRMP stars are \citet{venn2012}, \citet{sakari2019}, and \citet{Reggiani2023}}
\end{center}
\end{figure*}

\begin{figure*}
\begin{center}
\includegraphics[scale=0.7]{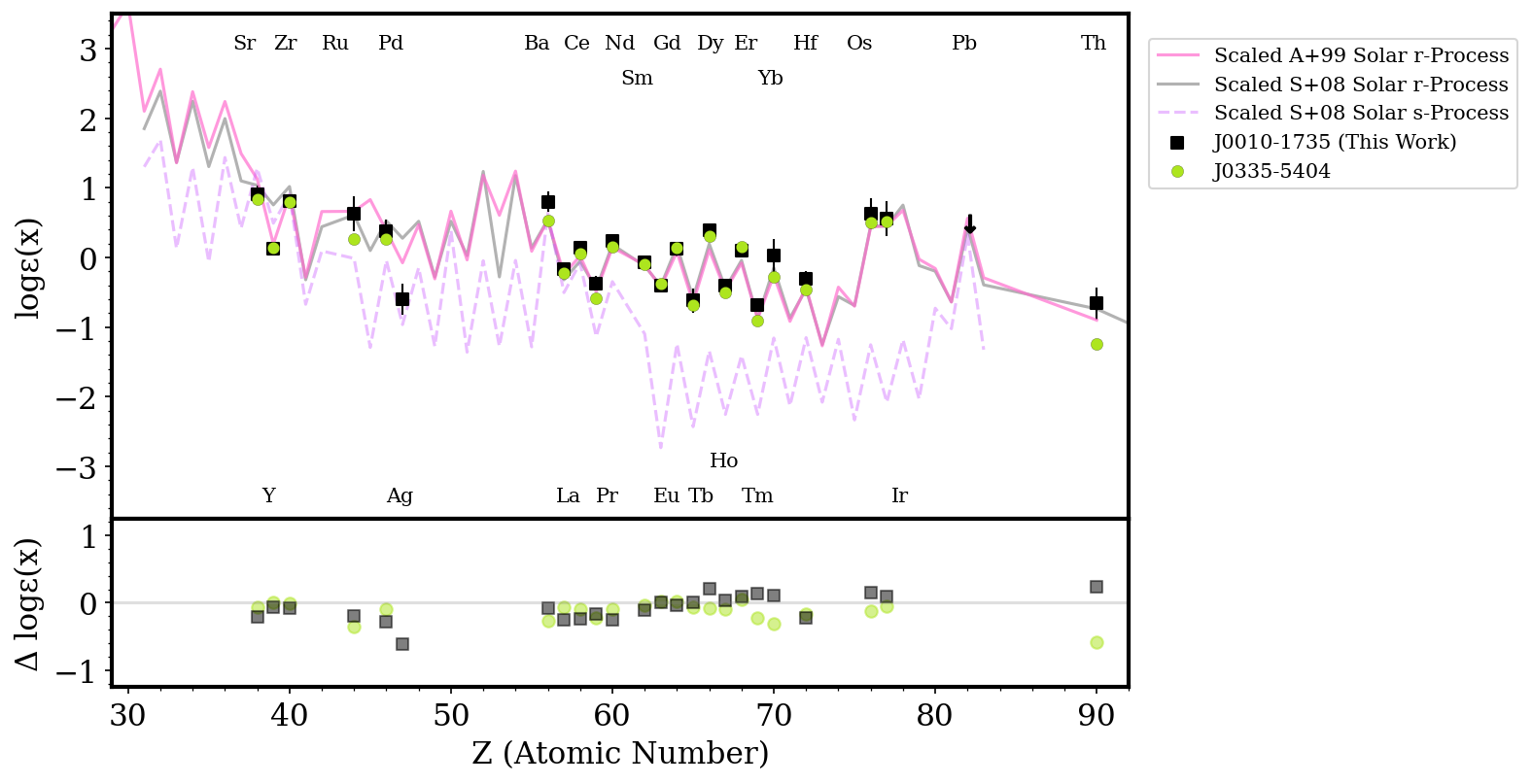}
\caption{\label{fig:rprocpatt} Elemental abundances of neutron-capture elements in \thestar\ (black squares), plotted over the \citet{Arlandini1999} Solar \rproc\ pattern (pink line), and \citet{sneden2008} Solar \rproc\ pattern (gray line), all scaled to \thestar's europium abundance of [Eu/Fe] = +1.30. The green circles depict abundances from the the brightest star in Reticulum II, J0335$-$5404 \citep{ji2016b}. The purple dashed line is the Solar \sproc\ \citep{sneden2008} pattern scaled to our star's Ba abundance, showcasing that our star does not have notable contributions from the \sproc. The bottom panel shows the residuals between our star and the scaled \citet{sneden2008} Solar abundance pattern (gray squares), as well as the residuals between \thestar\ and the Ret II star J0335$-$5404 (green circles).}
\end{center}
\end{figure*}

We derive Fe-peak abundances for \scii, \tii, \tiii, \scii, \vi, \vii, \cri, \crii, \mni, \mnii, \coi, \nii, and \zni. Abundances for \tii\ and \tiii\ were calculated using 18 and 29 lines, respectively, yielding [\tii/Fe] = $-0.19$ and [\tiii/Fe] = $-0.18$. Sc, \vi, and \vii, were derived from eight, one, and four lines, respectively, yielding [Sc/Fe] = $-0.87$, [\vi/Fe] = $-0.90$, and [\vii/Fe] = $-0.70$. Cr was detected from 10 \cri\ and 4 \crii\ lines, giving [\cri/Fe] = $-0.19$ and [\crii/Fe] = $-0.06$. \mni\ and \mnii\ were detected with eight and three lines, yielding [\mni/Fe] = $-0.77$ and [\mnii/Fe] = $-0.64$. The \coi\ abundance was estimated using 20 lines, resulting in [Co/Fe] = $-0.20$. We derived [Ni/Fe]$ = -0.48$ from 26 \nii\ lines. The Zn abundance was derived from the single detected \zni\ line at 4810\,{\AA}, leading to [Zn/Fe] $ =-0.46$.

On average, the Fe-peak abundances for \thestar\ are on the lower end of the abundances expected for a typical MW halo star, as shown in Figure~\ref{fig:abundplot}. We note that the \vi\ and \vii\ abundances lie well below that of typical stars in the [X/Fe] plane, yet all of the V lines yield abundances that agree within 0.1 dex of each other. 

For the Fe-peak elements, we applied NLTE-corrections to the abundances of \tii, \tiii, \mni, \coi\ and \nii. \coi\ exhibited the largest departure between NLTE and LTE values, with [Co/Fe]$_{\text{NLTE}} $ = +0.60, nearly 0.8 dex higher than the LTE abundance. The NLTE abundances for the Fe-peak elements are [\tii/Fe]$_{\text{NLTE}} $ =$+$0.05, [\tiii/Fe]$_{\text{NLTE}} $ = $-0.09$, [\mni/Fe]$_{\text{NLTE}} $ = $-0.49$, [Co/Fe]$_{\text{NLTE}}$ = $+$0.38, and [Ni/Fe]$_{\text{NLTE}} $= $-0.28$. However, as NLTE corrections are not available for all Fe-peak elements in our line list, we adopt the LTE abundances throughout our analysis for consistency.

\subsubsection{Neutron-capture Elements}\label{sec:ncap_abund}


We show the spectral regions around representative heavy-element lines in Figure~\ref{fig:spectrum}. We derived abundances and upper limits for 28 neutron-capture elements in \thestar, from Sr to Th. Three \srii\ lines were used to determine [Sr/Fe]= +0.26. Barium abundances were derived from five \baii\ lines yielding [Ba/Fe] = $+0.85$. Eight europium lines were used to derive an average europium abundance of [Eu/Fe] = $+1.30$. We use the Sr, Ba and Eu abundances to classify \thestar\ as a highly \rproc\,-enhanced \rII\,  star following \citet{beers2005} and \citet{rpa4}, where \rII\ stars are classified as [Eu/Fe]$>+1.0$ and [Ba/Eu]$<0.0$. The ratio [Ba/Eu] = $-0.45$ indicates that the neutron-capture abundance pattern of \thestar\ is dominated by \rproc\ nucleosynthesis rather than the \sproc\ \citep[e.g.,][]{sneden2008}. 

In addition to Sr, Ba, and Eu, we derived abundances for \ion{Y}{II}, \ion{Zr}{II}, \ion{Mo}{I}, \ion{Ru}{I}, \ion{Rh}{I}, \ion{Pd}{I}, \ion{Ag}{I}, \ion{La}{II}, \ion{Ce}{II}, \ion{Pr}{II}, \ion{Nd}{II}, \ion{Sm}{II}, \ion{Gd}{II}, \ion{Tb}{II}, \ion{Dy}{II}, \ion{Ho}{II}, \ion{Er}{II}, \ion{Tm}{II}, \ion{Yb}{II}, \ion{Lu}{II}, \ion{Hf}{II}, \ion{Os}{I}, and \ion{Ir}{I}, as well as report an upper limit for \ion{Pb}{I} of [\ion{Pb}{I}/Fe]$\leq$0.96. Average abundances are listed in Table~\ref{tab:chemabun}. One actinide element, thorium, was detected in \thestar\, from the line at 4019\,{\AA}. Our heavy element abundances are in good agreement ($\Delta$[X/Fe]$\lesssim0.2$) with the neutron-capture abundances reported for \thestar\ by \citet{rpa3} and \citet{gull2021}. 

Figure~\ref{fig:rprocpatt} shows the neutron-capture element abundance pattern overlaid with the scaled Solar \rproc-abundance patterns from \citet{Arlandini1999} and \citet{sneden2008},  scaled to the Eu abundance of \thestar. 
For comparison, we also show the scaled \sproc\ Solar pattern from \citet{sneden2008} on the same figure, normalized relative to \thestar's Ba abundance.  Most abundances of \thestar\ match the Solar \rproc\ pattern well. Pb is reported as an upper limit and is represented by a down arrow.

The residuals of \thestar\ and the Solar \rproc\ pattern are shown in the bottom panel of Figure~\ref{fig:rprocpatt}, denoted by gray squares. For elements with only one line, we adopted an error floor of 0.2 dex to reflect the uncertainty. We also show the scaled abundance pattern of J0335$-$5404 (green circles), the brightest star in the $r$-process-enhanced UFD Reticulum II \citep{ji2018} to illustrate an example of an \rproc-enhanced UFD star. The green circles in the bottom panel show the residuals between the abundances of our star and that of the Reticulum II star. Most elements show excellent relative abundance agreement within $0.2$\,dex.

\begin{figure*}
\begin{center}
\includegraphics[scale=0.32]{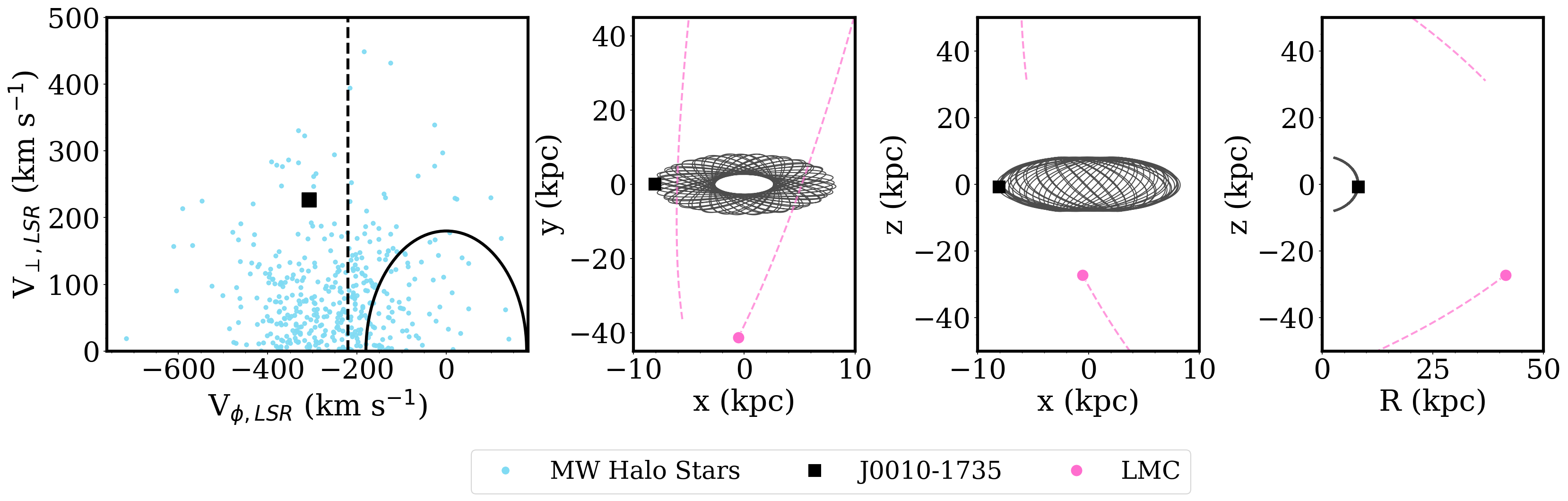}
\caption{\label{fig:action} Left: Toomre Diagram with \thestar\ (black square) and published RPA stars. The dashed black circle is a 180 \kms\ radius from the approximate location of the Local Standard of Rest, centered at V$_{\phi}$=0 \kms. The vertical black dotted line separates stars with prograde ($V_{\phi}>-220$ \kms) and retrograde ($V_{\phi}<-220$ \kms) orbits. Final three panels: The 8 Gyr backwards-integrated Galactocentric orbits in X, Y, and Z for \thestar\ (black) and the LMC (pink).}
\end{center}
\end{figure*}

\subsection{Kinematic Properties of \thestar}\label{sec:kinematics}

\citet{gull2021} included \thestar\ as part of their $\omega$\,Cen progenitor stream stars; however, in their discussion, they flagged it as having unusual abundances relative to other stars in their sample. This is also true with respect to \rproc\ elements, where \thestar\ is the only highly \rproc\ star identified in their $\omega$\,Cen sample. We therefore use the combination of chemistry and kinematics of \thestar\ in the following sections to discuss the possible origins of its chemical enrichment and assess its origin, whether from the MW halo or a satellite of the MW.  

The kinematic properties of \thestar\ are presented in Figure~\ref{fig:action}. The Toomre diagram in the leftmost panel displays the perpendicular velocity (V$_{\perp}$) combining the radial and vertical components as a function of the azimuthal velocity (V$_{\phi}$) for \thestar\ and other RPA stars \citep{rpa1,rpa2,rpa3,rpa4}. These velocities were derived from radial velocities and proper motions in RA and DEC from $Gaia$ DR3 \citep{gaiaedr3}. The vertical dashed line separates stars with prograde ($V_{\phi}>-220$) versus retrograde ($V_{\phi}<-220$) orbits. The black circle marks a 180 \kms\ radius centered on the Sun's position in the LSR ($V_{\phi,\odot}= 0$ \kms), such that stars within this boundary are kinematically consistent with MW disk membership, while those beyond it are more likely associated with the MW halo \citep[e.g.,][]{Venn2004,Nissen2010,Bensby2014,chen2021}. 
The Toomre diagram shows that \thestar\ is likely a member of the accreted halo due to its retrograde orbit (V$_{\phi,LSR}=-307$ \kms, V$_{\perp,LSR}$=226 \kms) \citep[e.g.,][]{SearleZinn1978,schlaufmann2012,bonaca2017,helmi2018}.

The final three panels of Figure~\ref{fig:action} show the 8 Gyr backwards-integrated orbits of \thestar\ in the Galactic X, Y, and Z planes using the galactic dynamics python package \texttt{Gala} \citep{gala,gala_software}. The mass-model of the MW used for the \texttt{gala.dynamics.orbit} function was the \texttt{MilkyWayPotential2022}, which has been fit to the rotation curve in \citet{Eilers2019} and incorporates the phase-space spiral in the solar neighborhood set by \citet{DarraghFord2023}. The orbit of \thestar\ has spherical apocenter and pericenter radii of $r_{apo}\approx 8.45$ kpc and $r_{peri}\approx 8.10$ kpc, corresponding to an eccentricity of $e\approx0.02$. The maximum height above the plane of the orbit is $z_{max}\approx7.97$ kpc. The LMC orbit is integrated backward in the same static potential using the present-day phase-space position and proper motions from \citet{vandermarel2014}. We note that this approach does not account for the gravitational influence of the LMC on the MW potential or dynamical friction, which would be required for a fully self-consistent treatment \citep{garavito2019, vasiliev2021}. We found that the qualitative comparison between the orbital morphology of \thestar\ and the LMC shows no dynamical association, effectively ruling out that \thestar\ was accreted by the LMC or an LMC satellite. 

\begin{figure*}
\begin{center}
\includegraphics[scale=0.45]{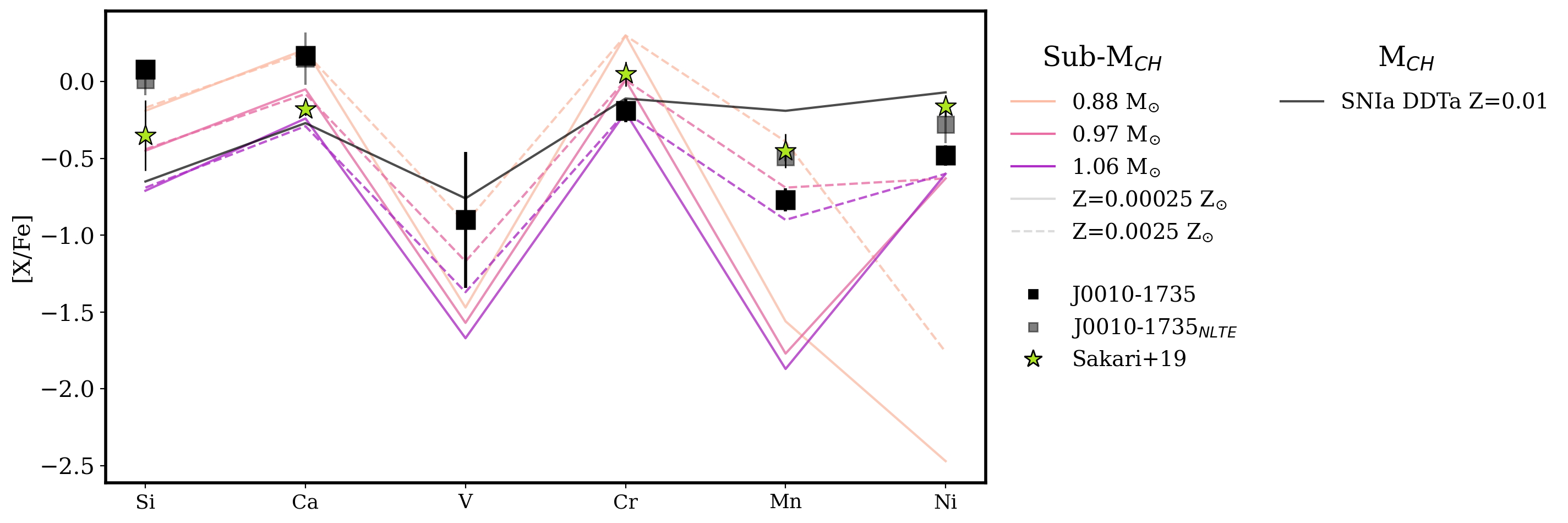}
\caption{\label{fig:yields} Comparisons of \thestar's abundances (black squares) and different yields from sub-\chandmass\ Ia SNe \citep[colored lines;][]{mcwilliam2018} and one \chandmass\ Ia SNe \citep[gray line;][]{Badenes2003,mcwilliam2018}. We overplot the IRMP star from \citet{sakari2019} as green star symbols. The NLTE-corrected abundances of \thestar, where available, are shown with gray squares.}
\end{center}
\end{figure*}

\section{Discussion}\label{sec:res}

\subsection{Chemical-enrichment Scenarios of \thestar}\label{sec:chemenrichment}

One of the most notable characteristics of \thestar\ is its sub-solar [X/Fe] ratios for most $\alpha$ elements, odd-Z light elements, and Fe-peak elements (Na, Mg, Al, Si, K, Sc, Ti, V, Cr, Mn, Co, Ni, and Zn) as shown in Figure \ref{fig:abundplot}. With [Eu/Fe]=+1.30, this is also a highly \rproc-enhanced $r$-II star as shown in Figure \ref{fig:rprocpatt}. We discuss the potential chemical-enrichment scenarios in the following sub-sections.

\subsubsection{Light, $\alpha$, and Fe-peak Enrichment}

Based on a smaller number of abundances, \cite{gull2021} suggested that this star belongs to a class of stars deemed ``iron-rich" metal-poor (IRMP) stars. While there is not yet a clear uniform enrichment channel for IRMP stars, they are considered to have formed from gas chemically enriched predominantly by a single (early) Type Ia supernova, that enhanced certain Fe-peak element abundances, while the abundances of other elements remain sub-solar with respect to Fe \citep[e.g.,][]{Ivans2003,venn2012,yong2013,Jacobson2015,mcwilliam2018,sakari2019}. In particular, \citet{Reggiani2023} defined IRMP stars as having [Na,Mg,Co/Fe] $<$ 0 and found that the progenitors of these stars are best described as double detonations of sub-Chandrasekhar (\chandmass) CO white dwarfs. \thestar\ satisfies this [Na,Mg,Co/Fe] $<$ 0 criterion, and we consider the IRMP classification as one plausible origin.

Our detailed chemical abundance analysis of \thestar\ enables the possibility of characterizing the Type Ia SNe progenitor. In particular, Type Ia SNe of sub-\chandmass\ white dwarfs would exhibit low Mn and Ni yields, whereas Type Ia SNe with \chandmass\ Ia progenitors would not \citep{mcwilliam2018,kirby2019,delosreyes2020,reggiani2021}. As \thestar\ exhibits sub-solar Mn and Ni abundance ratios ([\mni/Fe] = $-$0.77; [Ni/Fe] = $-$0.48), we compare the Fe-peak abundances of \thestar\ to theoretical yields of sub-\chandmass\ and near-\chandmass\ Type Ia SNe.

Figure~\ref{fig:yields} shows the LTE (and wherever available NLTE) abundances of \thestar\ (as well as the RPA IRMP star from \citealt{sakari2019}) with different sub-\chandmass\ Ia SN yields. The sub-\chandmass\ yields from \citet{mcwilliam2018} differ in their nucleosynthetic yields from those of \chandmass\ Ia SNe \citep{Badenes2003,mcwilliam2018} due to a neutron excess occuring in \chandmass\ progenitors. With a $\chi^2$ minimization, we found that the best-fit yield model to \thestar's LTE abundances is the sub-\chandmass\ Ia yield with a progenitor that is 0.97 M$_{\odot}$ and 0.0025 Z$_{\odot}$. The best-fitting yield to the NLTE abundances is a progenitor that is 0.88 M$_{\odot}$ and 0.0025 Z$_{\odot}$.  

We note, however, that the sub-solar [Mn/Fe] and [Ni/Fe] ratios are also a generic signature of extended, Ia-dominated chemical evolution at metallicities of [Fe/H] $> -1$ of UFDs \cite[e.g.,][]{Venn2004,kirby2019}, rather than evidence unique to an IRMP star enriched by a single sub-\chandmass\ event. However, as \thestar\ has a metallicity of [Fe/H] = $-2.22$, this is unlikely the case for this star.

Another chemical peculiarity of \thestar\ is the high Ca abundance ([Ca/Fe] = $+0.17$) relative to the other $\alpha$-element abundance ratios. Sub-\chandmass\ SNe Ia are predicted to produce higher fractions of Ca compared to \chandmass\ SNe Ia, although this would also produce higher amounts of Si, S, and Ar \citep{Kobayashi2020}, which is not the case in \thestar\ ([Si/Fe] = $+0.08$). Additionally, \citet{Chen2026} found that high-redshift SNe Ia contain more Ca in their ejecta than their low-redshift counterparts, which could potentially support the conclusions in Figure \ref{fig:yields}. 

Beyond typical CCSNe and SNe Ia, \citet{perets2010} identified a class of explosions termed ``Ca-rich supernovae" based on their high Ca/O nebular line ratios and intermediate luminosities between novae and standard SNe, later dubbed ``Ca-rich gap transients" by \citet{kasliwal2012}. Relevant to \thestar, \citet{kirby2019} found that [Ca/Fe] declines more slowly than [Si/Fe] in dSphs, resulting in significantly higher inferred yields of [Ca/Fe]$_{\text{Ia}}$ than [Si/Fe]$_{\text{Ia}}$ from SNe Ia. While this behavior is predicted by sub-\chandmass\ models \citep{blondin2018,Shen2018}, an alternative explanation is that Ca-rich gap transients contribute an additional delayed source of Ca enrichment, as these events are suggested to produce substantial Ca while yielding comparatively little Fe-peak material. We note that this scenario hinges on the non-detection of oxygen, and if [O/Fe] was detected, a Ca-rich transient would be ruled out.

Taking all these lines of evidence, we thus postulate that the overall abundance pattern of \thestar\ could be explained by enrichment from sub-\chandmass\ Ia SNe, potentially supplemented by an additional Ca-rich gap transient. The sub-solar [Mn/Fe] and [Ni/Fe] ratios are consistent with the Fe-peak yields characteristic of sub-\chandmass\ detonations \citep{mcwilliam2018,kirby2019,delosreyes2020}, while the star's [Ca/Fe] relative to its other $\alpha$-elements suggests an additional source of Ca beyond what sub-\chandmass\ models alone predict. 

\subsection{\rproc\ Enrichment}

In addition to these sources, a separate, neutron-dense enrichment site is required to produce the observed \rproc\ enhancement ([Eu/Fe]=+1.30). The \rproc\ pattern for \thestar\ follows the universal pattern that has been identified in previous studies \citep[i.e.,][]{roederer2022}, hinting at a similar underlying astrophysical site. Likely sources for \rproc-enrichment include neutron star mergers \citep{abbott2017,drout2017}, jet-driven magnetorotational supernovae \citep{nishimura2017}, or collapsars \citep{siegel2019}, although the latter two would likely also produce $\alpha$ and Fe-peak elements \citep[e.g.,][]{Zha2026}.

This extreme level of \rproc-enrichment closely mimics the chemistry of Reticulum II (compared in Figure \ref{fig:rprocpatt}), which is believed to have been polluted by a single, prolific, neutron star merger event. The similarity to Reticulum II supports a similar enrichment history such as a compact binary merger as the source for \rproc-enhancement in this star. 

\subsection{Birth Environments}\label{sec:environments}

The chemical abundance ratios of \thestar\ are systematically sub-solar and deficient in nearly all light, $\alpha$, and Fe-peak elements relative to typical MW field halo stars \citep[Figure~\ref{fig:abundplot};][]{roederer2014, yong2013}, with the notable exception of Ca. This systematic deficiency motivates an investigation into the origin of \thestar, and whether it is truly a native MW halo star or if it has \textit{ex-situ} origins.

\subsubsection{Comparisons with MW Halo Stars}\label{sec:mwstars}

\begin{figure}
\begin{center}
\includegraphics[scale=0.55]{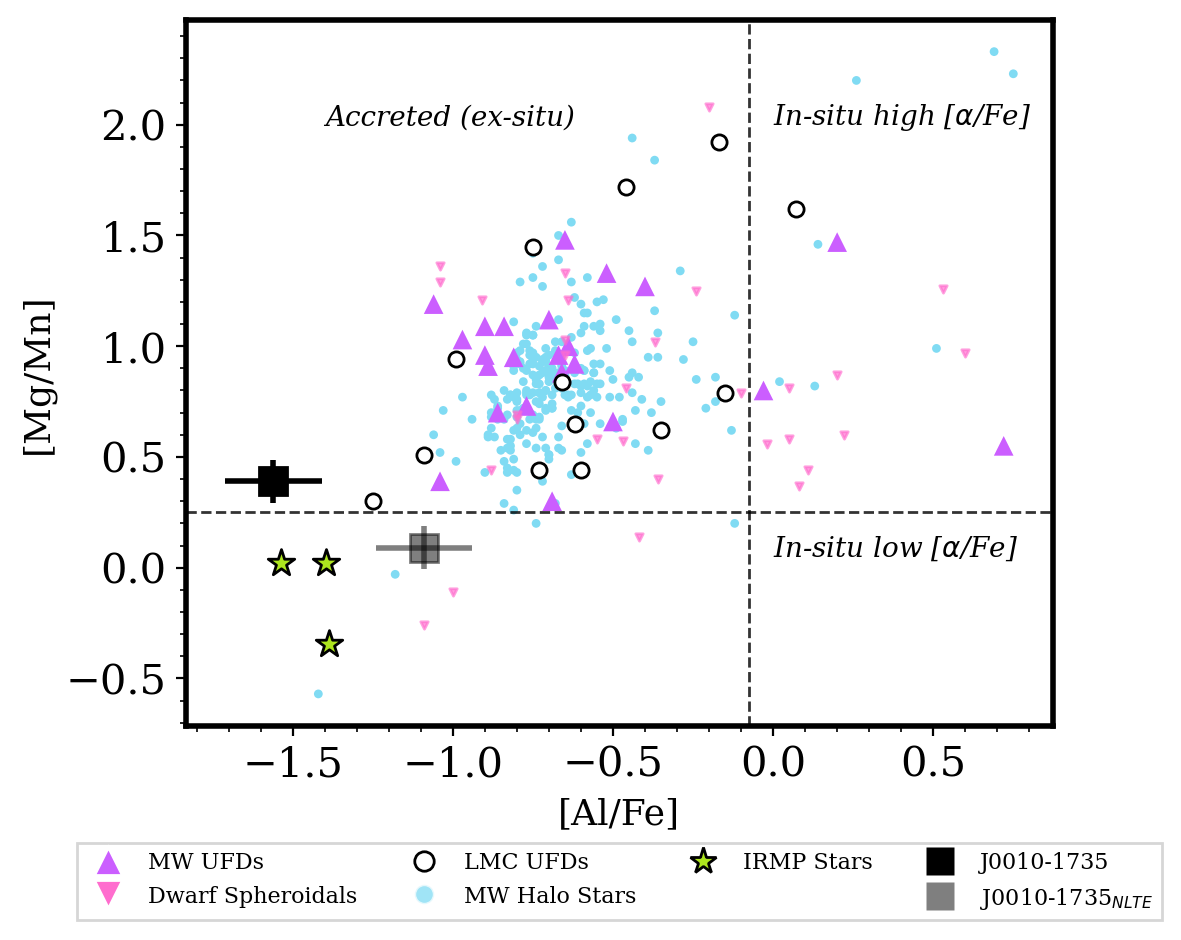}
\caption{\label{fig:mgmn} [Mg/Mn]--[Al/Fe] plane that chemically separates $in$-$situ$ and $ex$-$situ$ components of the MW due to different chemical-enrichment and star-formation histories \citep{hawkins2015,das2020,belokurov2022,marasco2025}. \thestar\ clearly falls in the accreted region, which supports an $ex$-$situ$ origin. Symbols are the same as in Figure~\ref{fig:abundplot}.}
\end{center}
\end{figure}

The [Mg/Mn]--[Al/Fe] plane has been established as a powerful diagnostic for distinguishing accreted (\textit{ex-situ}) from \textit{in-situ} stellar populations, independent of kinematics \citep{hawkins2015,das2020,horta2021,belokurov2022,marasco2025}. This plane combines the enrichment channels of core-collapse supernovae (CCSNe) versus Type Ia SNe traced by [Mg/Mn] \citep[e.g.,][]{Tinsley1980,Timmes1995,Thielemann1996,Matteucci2001,Kobayashi2006,Simon2019rev,Arcones2023}, and the metallicity-dependent nucleosynthetic enrichment of Al \citep[e.g.,][]{Woosley1995,Kobayashi2020,hawkins2015}, which differs significantly between stars born in the early MW and those accreted from lower-mass systems with lower star formation rates. Figure~\ref{fig:mgmn} shows \thestar\ in this plane alongside MW halo stars, stars from UFDs, stars from LMC UFDs, classical dwarf spheroidal galaxy stars, and other IRMP stars. \thestar\ falls firmly in the accreted region, with [Al/Fe] = 
$-$1.55 and [Mg/Mn] = $+0.39$, consistent with an origin in an accreted system rather than the \textit{in-situ} MW halo, which is likely the case of a majority of MW halo stars \citep[e.g.,][]{Mackereth2019}.

\subsubsection{Comparisons with Satellite System Stars}\label{sec:mwsatellite}

\begin{figure*}
\begin{center}
\includegraphics[scale=0.8]{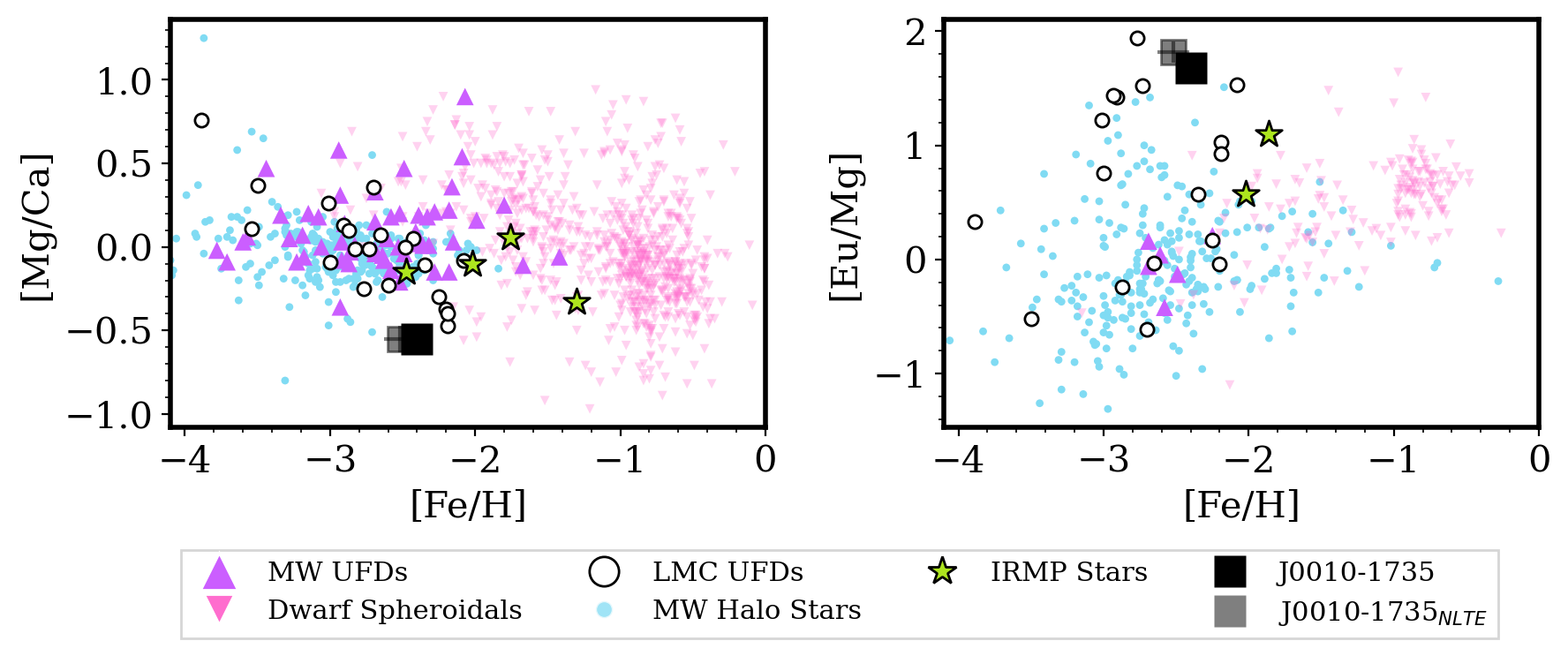}
\caption{\label{fig:eumg} [Mg/Ca] and [Eu/Mg] ratios, with respect to the metallicity of \thestar\ (black square) and other systems, with the same symbols as in Figure~\ref{fig:abundplot}. The green star is the IRMP star J0937$-$0626 \citep{sakari2019}. These abundance ratios trace the chemical-enrichment histories of different types of satellite systems.}
\end{center}
\end{figure*}

As the chemistry and kinematics of \thestar\ point towards \textit{ex-situ} accretion origins, we explore potential progenitor systems. Globular clusters (GCs) are well known for their characteristic light element anti-correlations, where a fraction of stars show strongly enhanced Al (and Na) at the expense of O (and Mg), produced by high-temperature proton-capture nucleosynthesis in the CNO and NeNa/MgAl cycles in an earlier generation of massive AGB stars or rapidly rotating massive stars \citep{carretta2009,bastian2018}. Halo stars with GC origins therefore show elevated [Al/Fe], whereas \thestar\ is depleted in Al with [Al/Fe] = $-1.55$ (shown in Figure~\ref{fig:mgmn}), which makes a GC origin for \thestar\ highly unlikely.

Dwarf spheroidal (dSph) (with luminosities of $> 10^5$ $L_{\odot}$) and UFD (luminosities of $ < 10^5$ $L_{\odot}$) satellite galaxies host some of the oldest, most metal-poor, and least chemically evolved stellar populations known \citep{Simon2019rev}. Given that the chemical-abundance pattern of \thestar\ is consistent with an accretion origin, likely an early analog of a surviving dwarf satellite, we explore its connection to stars in current dwarf galaxies.

Figure~\ref{fig:eumg} shows different abundance ratios with respect to metallicity to highlight the various relative chemical-enrichment histories of these systems. [Mg/Ca] in the left panel traces the relative contributions of different types of CCSNe. Mg is produced predominantly in massive, short-lived stars via hydrostatic carbon and neon burning, while Ca is produced in both CCSNe and Type Ia supernovae; however, it also has contributions from a broader range of progenitor masses, including lower-mass CCSNe \citep{Woosley1995, Kobayashi2006, Kobayashi2020}. A sub-solar [Mg/Ca] ratio suggests either a relative deficiency in the most massive stellar progenitors or an enhanced contribution from lower-mass CCSNe (discussed further in Section \ref{sec:chemenrichment}). 

The right panel of Figure~\ref{fig:eumg} shows the production ratio of Eu to Mg. Eu is an \rproc\ element thought to be produced by neutron star mergers \citep{abbott2017,drout2017}, jet-driven magnetorotational supernovae \citep{nishimura2017}, or collapsars \citep{siegel2019}, while Mg is produced largely by CCSNe. An elevated [Eu/Mg] therefore indicates that the system experienced significant \rproc\ enrichment relative to its $\alpha$-element content. In low-mass host systems, it is indeed possible that a single rare \rproc\ event contributes heavy elements across the entire galaxy \citep[i.e.,][]{ji2016Nat}.

In dSph galaxies, an extended duration of star formation allows Ia SNe to produce Fe early on, shifting the ``$\alpha$-knee" to lower metallicities, pushing all [$\alpha$/Fe] abundances to sub-solar levels \citep{Venn2004}. As described above, this is not found for Ca and Si of this star. The sub-solar [Mn/Fe] and [Ni/Fe] ratios are consistent with enrichment by sub-\chandmass\ SNe Ia progenitors in a system with an early, truncated star formation history \citep{seitenzahl2013,kirby2019}. \thestar's overall [X/Fe] abundance pattern also matches the MW UFD population (Figure~\ref{fig:abundplot}), and the star shows strong \rproc\ enhancement. Taken together, we consider the likely progenitor system for \thestar\ to be an accreted early analog of a UFD, plausibly one with an enrichment history simar to that of Reticulum II.

We note that this interpretation, driven by ordinary SNe Ia-dominated chemical evolution within a low-mass dwarf galaxy, provides an alternative to the single-Ia SNe IRMP classification as discussed in Section \ref{sec:chemenrichment}. 


Several satellite UFDs, such as Reticulum II, Horologium I, Carina II and Carina III, have recently been identified as being associated with the Large and Small Magellanic Cloud system, the most massive satellites of the MW \citep{kallivayalil2013,kallivayalil2018,erkal2020,ji2026}. However, as \citet{ji2020car} notes and this present study reinforces, the chemical distinction between MW UFDs and LMC satellite UFDs is not yet sufficiently statistically significant to draw strong conclusions. Solely based on the kinematics presented in Section \ref{sec:kinematics} and Figure~\ref{fig:action}, we conclude that LMC satellite origins are unlikely for \thestar. 

\subsection{The Age of \thestar\ and its Host System}\label{sec:age}

The detection of the actinide element Th in \thestar\ provides an opportunity to apply nuclear cosmo-chronometry, exploiting the radioactive decay of Th on cosmological timescales to place an age constraint on the star \citep{Cowan1991,Sneden2002,placco2017}. Given the demonstrated $ex$-$situ$ origin of \thestar, this age estimate also serves as a lower limit on the age of the progenitor UFD-analog from which it was accreted. 

The most precise chronometer to derive ages from would be U/Th, as these are both actinide elements and would have the most similar formation conditions \citep{Cowan1991,Sneden2003,Frebel2007_u,Hill2017,shah2023,Huang2025}. However, the strongest uranium line at 3859\,{\AA} in \thestar\ is too weak to yield a reliable U detection. We therefore rely primarily on the [Th/Eu] ratio, which has been well-established as an age estimate for \rproc-enhanced stars \citep[e.g.,][]{cain2018}. 
We caution, however,  that the Th/Eu chronometer has been shown to be highly sensitive to abundance uncertainties, the chosen production ratio, and the principal choice for the \rproc\ production site \citep{Hill2017,holmbeck2018,azhari2025}. In light of that, ages should be carefully treated as their uncertainties are significant.

Following the methodology of \citet{placco2017}, we adopt the production ratio from high entropy wind \rproc\ models by \citet{hill2016}, as these production ratios include updated nuclear physics models and reproduce the universal \rproc\ pattern. We then use the following equation from \citet{schatz2002}:


\begin{equation}
    \tau = 46.7\,  \text{Gyr}\,   [\text{log(Th/X)}\mathrm{o} - \text{log(Th/X)}\mathrm{*}]
\end{equation}
\noindent where X is a stable \rproc\ element, Eu in this case. $\text{Log(Th/X)}_o$ corresponds to the initial production ratio at time = 0, and $\text{log(Th/X)}_*$ corresponds to the abundance ratio measured in the star today. 


At face value, this delivers an age with the Th/Eu chronometer ($\log\epsilon$(Th/Eu)$=-0.26\pm0.24$) of 12.5 $\pm$ 4.2 Gyr. This age indicates the progenitor system formed its stars very early in cosmic history, consistent with a rapid enrichment timescale of an UFD.

\section{Summary and Conclusions}\label{sec:conc}

We report a detailed chemo-dynamical analysis of \thestar, a metal-poor ([Fe/H] = $-2.22$) highly \rproc\ enhanced ([Eu/Fe] = +1.30) star located in the MW Halo. We derived over 40 elemental abundances, including light, $\alpha$, Fe-peak, and neutron-capture elements for the star. Investigation of its chemical abundances shows that \thestar\ exhibits significant sub-solar [X/Fe] abundance ratios in nearly all of its light, $\alpha$, and Fe-peak elements relative to the MW halo, with the exception of Ca.

The chemical-abundance pattern of the light, $\alpha$, and Fe-peak elements of this star could suggest that it is an ``iron-rich metal-poor" star \citep[e.g.,][]{venn2012,yong2013,mcwilliam2018,sakari2019,reggiani2021}, or simply a star with a similar chemical-enrichment history to Reticulum II. The sub-solar [Mn/Fe] and [Ni/Fe] ratios are consistent with enrichment of the birth gas cloud by sub-\chandmass\ SNe Ia progenitors in a system with an early, truncated star-formation history \citep{seitenzahl2013,kirby2019}. The elevated [Ca/Fe] ratio relative to that of other $\alpha$-elements may additionally reflect a contribution from Ca-rich gap transients \citep{perets2010, kasliwal2012}. 

The neutron-capture abundance pattern from Sr to Th closely follows the scaled solar \rproc\ residual pattern \citep{sneden2008, Arlandini1999}, with [Ba/Eu] = $-0.45$ confirming that the neutron-capture content of \thestar\ is dominated by \rproc\ nucleosynthesis rather than the \sproc. This is consistent with enrichment from a single or small number of \rproc\ events in a low-mass progenitor system, where the stochastic nature of rare nucleosynthetic events can dominate the chemical budget of the entire galaxy \citep{ji2016Nat,roederer2018}.

We assess the potential progenitor system of \thestar\ by comparing its properties to those expected for MW halo and satellite galaxy populations. This star lies firmly in the accreted region of the [Mg/Mn]--[Al/Fe] plane \citep[i.e.,][]{hawkins2015}, indicating that it was likely accreted from an early satellite system. Combined with a retrograde orbit and an abundance pattern resembling those of MW satellite stars, this supports an \textit{ex-situ} origin for \thestar.
We compare the kinematics of \thestar\ to the LMC and found that this star has a high-inclination polar orbit confined to the inner halo, distinct from the trajectory of the LMC system, making an LMC satellite origin unlikely. The chemical abundances of this star, namely the low Fe-peak abundances combined with the high Ca abundance, point towards \thestar\ originating in an early analog of a surviving UFD satellite. 

As large-scale surveys such as the RPA \citep{rpa1, rpa2, rpa3, rpa4, rpa5} continue to expand the census of chemically characterized metal-poor stars, the identification of accreted stars like \thestar\ will place increasingly stringent constraints on both the nucleosynthetic origins of the \rproc\ and the disrupted satellite systems that built the MW halo. 


\facilities{Magellan-Clay (MIKE, \citealt{bernstein2003})}

\software{\texttt{IRAF} \citep{iraf,fitzpatrick2025}, \texttt{MOOG} \citep{sneden1973,sobeck2011}, \texttt{LOTUS} \citep{lotus}, \texttt{MULTI} \citep{Carlsson1986,carlsson1992}, \texttt{Turbospectrum} \citep{alvarez1998, plez2012, gerber2023, plez2025}, \texttt{TSFitPy} \citep{y_nlte}, \texttt{astropy} \citep{astropy:2022,astropy:2013}, \texttt{matplotlib.pyplot} \citep{matplotlib}, \texttt{numpy} \citep{numpy}, \texttt{scipy} \citep{scipy2020}, \texttt{Gala} \citep{gala,gala_software}}

\section*{Acknowledgments}
Z.H. thanks Keith Hawkins for fruitful Socratic discussions about the chemodynamics of stars. Z.H. thanks Patrick Hackshaw for nurturing her curiosity about the formation of the Universe. 

Z.H. is partially supported by NSF AST2407975. R.E. acknowledges support from NASA Astrophysics Theory Program grant 80NSSC24K089. R.E. and R.S. acknowledge support from the Future Investigators in NASA Earth and Space Science and Technology (FINESST) grant 24-ASTRO24-0129, as well as from the University of Florida Research Computing (UFRC) for providing computational resources that contributed to the research results reported in this publication. T.C.B. acknowledges partial support from grants PHY 14-30152; Physics Frontier Center/JINA Center for the Evolution of the Elements (JINA-CEE), and OISE-1927130; The International Research Network for Nuclear Astrophysics (IReNA), awarded by the US National Science Foundation, and DE-SC0023128; the Center for Nuclear Astrophysics Across Messengers (CeNAM), awarded by the U.S. Department of Energy, Office of Science, Office of Nuclear Physics. The work of V.M.P. is supported by NOIRLab, which is managed by the Association of Universities for Research in Astronomy (AURA) under a cooperative agreement with the U.S. National Science Foundation. T.T.H. acknowledges support from the Swedish Research Council (VR 2025-05377) and the Olle Engkvist foundation. A.F. acknowledges support from NSF-AAG grant AST-2307436.

\bibliography{ref}{}
\bibliographystyle{aasjournalv7_update}

\clearpage
\appendix

\begin{deluxetable}{lrrrrrr}[h!]
\tabletypesize{\normalsize}
\tablecaption{Chemical Line Measurements of \thestar\label{tab:linemeasurements}}
\tablehead{
\colhead{$\lambda$} & \colhead{Species} & \colhead{$\chi$} & \colhead{$\log gf$} & \colhead{EW} &  \colhead{$\log\epsilon$(X)} &  \colhead{Ref.} \\
\colhead{[\AA]} & \colhead{} & \colhead{[eV]} & \colhead{} & \colhead{[m\AA]}  & \colhead{[dex]} & \colhead{}
}
\startdata
    5895.920 &   Na I &  0.000 & $-$0.190 &  80.03    &  3.32 & 2 \\
    4057.510 &   Mg I &  4.350 & $-$0.900 &  13.42    &  4.98 & 2 \\
    4167.270 &   Mg I &  4.350 & $-$0.740 &  22.62    &  5.09 & 2 \\
    4702.990 &   Mg I &  4.330 & $-$0.440 &  30.43    &  4.91 & 2 \\
    5172.680 &   Mg I &  2.710 & $-$0.360 & 135.83    &  4.92 & 2 \\
    5183.600 &   Mg I &  2.720 & $-$0.170 & 155.48    &  4.95 & 2 \\
    5528.400 &   Mg I &  4.350 & $-$0.550 &  30.28    &  5.02 & 3 \\
    5711.090 &   Mg I &  4.350 & $-$1.840 &   2.57    &  5.10 & 3 \\
    3944.000 &   Al I &  0.000 & $-$0.640 & \nodata\  &  2.69 & 2 \\
    3961.520 &   Al I &  0.010 & $-$0.330 &  73.54    &  2.67 & 2 \\
    5948.540 &   Si I &  5.080 & $-$1.230 &   3.66    &  5.38 & 2 \\
    7742.720 &   Si I &  6.210 & $-$0.280 &   2.61    &  5.37 & 2 \\
    7698.960 &    K I &  0.000 & $-$0.180 &   6.57    &  2.42 & 2 \\
    4092.630 &   Ca I &  2.520 & $-$0.860 &   4.94    &  4.22 & 2 \\
    4283.010 &   Ca I &  1.890 & $-$0.220 &  52.78    &  4.35 & 2 \\
\enddata
\end{deluxetable}
Table \ref{tab:linemeasurements} includes the information for each elemental absorption line detected in \thestar, including the atomic line data (wavelengths, species, excitation potentials, oscillator strengths), equivalent widths (where available), and absolute abundances as determined under LTE. Atomic line data were adopted from \citet{roederer2018a}, which used \texttt{linemake} \citep{linemake} and the National Institute of Standards and Technology (NIST) Atomic Spectral Database \citep{kramida2013}. The full machine-readable table is available online. The individual references are listed in the table and and correspond to: (1) \citet{masseron2014}, (2) \citet{kramida2013}, (3) \citet{pehlivanrhodin2017}, (4) \citet{lawler1989,kurucz1995},(5) \citet{lawler2013}, (6) \citet{wood2013}, (7) \citet{lawler2014}, (8) \citet{wood2014_v}, (9) \citet{sobeck2007}, (10) \citet{lawler2017}, (11) \citet{denharthog2011}, (12) \citet{obrian1991}, (13) \citet{denhartog2014}, (14) \citet{belmonte2017}, (15) \citet{ruffoni2014}, (16) \citet{lawler2015}, (17) \citet{wood2014}, (18) \citet{roederer2012_line}, (19) \citet{biemont2011}, (20) \citet{ljung2006}, (21) \citet{mcwilliam1998}, (22) \citet{lawler2001,ivans2006}, (23) \citet{lawler2009}, (24) \citet{li2007,sneden2009}, (25) \citet{ivarsson2001,sneden2009}, (26) \citet{denhartog2003,roederer2008}, (27) \citet{lawler2006,roederer2008}, (28) \citet{lawler2001_eu,ivans2006}, (29) \citet{denharthog2006}, (30) \citet{lawler2001_tb}, (31) \citet{wickliffe2000}, (32) \citet{lawler2004,sneden2009}, (33) \citet{lawler2008}, (34) \citet{wickliffe1997, sneden2009}, (35) \citet{sneden2009}, (36) \citet{lawler2009}, (37) \citet{lawler2007}, (38) \citet{quinet2006}, (39) \citet{xu2007,cowan2005}, (40) \citet{biemont2000,roederer2012_line}, and (41) \citet{Nilsson2002}

\end{document}